\documentclass[11pt,twoside,a4paper]{article}
\renewcommand{\baselinestretch}{1.5}
\usepackage[utf8]{inputenc}
\usepackage[english]{babel}
\usepackage{geometry}
\usepackage{graphicx}
\usepackage{dsfont}
\usepackage{charter}
\usepackage{booktabs}
\usepackage[flushleft]{threeparttable}
\usepackage{lscape}
\usepackage[caption = false]{subfig}
\usepackage{dcolumn}
\newcolumntype{d}[1]{D{.}{.}{#1}}
\newcolumntype{B}[3]{>{\boldmath\DC@{#1}{#2}{#3}}c<{\DC@end}}
\newcommand\mc[1]{\multicolumn{1}{c}{#1}} % handy shortcut macro
\newcommand*{\thead}[1]{\multicolumn{1}{c}{\bfseries #1}}
\usepackage{bm} %% caption parameters
\usepackage[bitstream-charter]{mathdesign}
\usepackage[T1]{fontenc}

\usepackage{natbib}
\usepackage{amsmath}
\usepackage{multicol}

\usepackage{url}

\title{Bayesian shrinkage in mixture of experts models:\\Identifying robust determinants of class membership\thanks{The author gratefully acknowledges the insightful guidance of Sylvia Frühwirth-Schnatter. Moreover, I would like to thank Florian Huber, Michael Pfarrhofer, Max Böck, Niko Hauzenberger, Mike Wögerer and Phil Grammel for helpful comments and support.}}

\author{Gregor Zens\thanks{gzens@wu.ac.at - Department of Economics, Vienna University of Economics and Business; Welthandelsplatz 1, 1020 Wien, Austria}
}

\begin{document}

\maketitle

\begin{abstract}
\noindent A method for implicit variable selection in mixture-of-experts frameworks is proposed. We introduce a prior structure where information is taken from a set of independent covariates. Robust class membership predictors are identified using a normal gamma prior. The resulting model setup is used in a finite mixture of Bernoulli distributions to find homogenous clusters of women in Mozambique based on their information sources on HIV. Fully Bayesian inference is carried out via the implementation of a Gibbs sampler.
\end{abstract}

\section{Introduction}

Modeling heterogeneity in datasets is a common problem in applied statistics. The task is to find underlying clusters of similar observations that can be used to describe the data. A widespread and known method to accomplish this is finite mixture modeling, where the main idea is to model a single probability distribution as the weighted sum of a finite number of mixture densities. This technique can be used for model based clustering as well as density estimation. Finite mixtures are widely used in different research fields -- a rather common application in marketing research is discussed in \citet{lenk2000bayesian}, who employ mixture modeling techniques to find clusters of customers with similar behaviour. Earlier references discussing marketing applications are \citet{allenby1995} and \citet{rossi1996}. \citet{lubrano2016} use mixtures to find homogenous groups in a study of the income distribution of the United Kingdom. However, the model family also extends to time series analysis naturally as shown in \citet{frohwirth2008model}. An applied example is the Markov mixture model that \citet{fruhwirth2012labor} use to model the earning dynamics in the Austrian labour market. For a comprehensive overview of mixture models and estimation strategies, see \citet{fruhwirth2006}.\\
The main contribution of this article lies, however, in a popular extension of the standard mixture framework. In the most basic Bayesian mixture models, prior class membership is modeled using the component weights, that is the relative size of the mixture clusters. Essentially, this means that the highest prior membership probability is assigned to the largest group in the population. This assumption implicitly claims that each observation has the same prior probability of belonging to a specific group, neglecting other observable characteristics of the data point. To make use of additional information, it is also possible to model mixture parameters as a function of external covariates. Such a specification usually allows for a richer interpretation of the model output and might permit a more holistic use of datasets. This modeling technique is usually referred to as a \textit{mixture-of-experts} (MOE). Despite the name originating in the machine learning literature\footnote{\citet{jacobs1991adaptive} calls the mixture components \textit{experts} and considers the mixture weights as \textit{gating networks} resulting in the now widely used nomenclature.}, mixture-of-experts models have a wide range of applications, similar to standard mixture models. \citet{gormley2008mixture} develop a MOE model for rank data and \citet{gormley2010mixture} use the framework to model network data. MOE models also apply to time series (\citealp{huerta2003time}; \citealp{fruhwirth2012labor}) and longitudinal data (\citealp{tang2016mixture}). Related models have been discussed under different labels for quite some time now, for instance switching regression models (\citealp{quandt1972new}) or concomitant variable latent-class models (\citealp{dayton1988concomitant}). For a comprehensive overview of mixture-of-experts models, refer to \citet{gor-fru:mod}.\\
This article focuses on a specific problem arising when dealing with mixture-of-experts models where covariates are included to model class membership\footnote{More general mixture-of-experts models also allow for variables to be included to model the the mixture component parameters, see \citet{gor-fru:mod}.}. There is severe model uncertainty regarding the relevant covariates to include to model prior class membership (as pointed out by \citealp{anderson2016}). Both estimated coefficients and class membership estimates might be sensitive to the particular choice of explanatory variables included in the cluster membership part of the model. One way to resolve this issue is to rerun the model using cross validation as a crude sensitivity analysis. However, the process of choosing which variables to include remains arbitrary. Thus, various approaches for variable selection in mixture frameworks have been discussed in literature, both tackling the questions of which variables to include in the class membership part of the model and which covariates enter the component parameter part of the model. See for instance the generalized smooth finite mixture model from \citet{VILLANI2012121}, linear cluster-weighted models (\citealp{ingrassia2014model}; generalized in \citealp{ingrassia2015generalized}) or the models for high dimensional mixture regressions in \citet{gupta2007variable} and \citet{devijver2015finite}.\\
We propose the use of a continuous shrinkage prior in latent class mixture modeling to isolate robust determinants of class membership to overcome this problem. More specifically, we specify a Normal-Gamma prior (\citealp{griffin2010inference}) and use the P\'{o}lya-Gamma sampler from \citet{polson2013bayesian} for computations to isolate important predictors of class membership. This results in more efficient shrinkage and improved performance when compared to related methods like standard stochastic search variable priors (\citealp{george1993variable}; introduced to latent class models in \citealp{ghosh2011}), especially when the number of possible predictors is large and/or the sample size is small. We illustrate our approach through simulation studies and an empirical example using Demographics and Health Survey (DHS) data from Mozambique.\\
The remainder of this work is organized as follows: Section 2 introduces the modeling framework. In Section 3, a simulation study is conducted to evaluate the performance of the proposed prior setup. Section 4 illustrates the framework in an application to HIV information source data from Mozambique. Section 5 concludes.

\section{Statistical Framework}
\label{sec:stats}
\subsection{Mixture-of-experts Models}

Let $y_i$ denote an observation of data point $i=1,\ldots,N$. This dependent variable can be univariate or multivariate, discrete or continuous, or of a more general structure such as time series or network data. Let $x_i$ be a set of $P$ (for $p=1,\ldots,P$) covariates of $y_i$. Assume $K$ (for $k=1,\ldots,K$) clusters exist in the population that follow known probability density functions $f_k(\cdot|\Xi_k)$ with component specific parameters $\Xi_k$. Denote the component weights as $\eta_k(x_i)$ where $\eta_k(x_i) \geqslant 0$ and $\sum_{k=1}^K \eta_k(x_i) = 1$. Then $y_i$ follows the mixture distribution

\begin{equation}
\label{MOE}
f(y_i ~|~ x_i) = \sum_{k=1}^K \ \eta_k(x_i) ~ f_k(y_i~|~\Xi_k).
\end{equation}

We assume the component weights $\eta_k(x_i)$ to be a function of the concomitant variables $x_i$. These covariates influence the distribution of $y_i$ indirectly via the individual prior class membership probabilities $Pr(S_i = k | x_i) = \eta_k(x_i)$. $S_i$ is the latent class membership indicator of individual $i$, where $S_i = k$ if $y_i$ belongs to cluster $k$. This relationship is typically modeled via a multinomial logit link with

\begin{equation}
\label{eq:mnlogit}
Pr(S_i = k | x_i, \beta_1,\ldots,\beta_{K-1}) = \frac{\text{exp}(x_i'\beta_k)}{1 + \sum_{k=1}^{K-1} \text{exp}(x_i'\beta_k)}
\end{equation}

where we set $\beta_K = 0$ to achieve identification of the model. This directly results in the interpretation of the coefficients in terms of a change in the log odds relative to the baseline category $K$. Other possibilities to model this ''gating function'' are discussed in \citet{yuksel2012twenty}.

\subsection{Prior Specification}

There are several ways to model Bayesian multinomial logistic regression. We choose the method proposed by \citet{polson2013bayesian} for simplicity and efficiency reasons. The Bayesian framework requires the specification of a prior on $\beta_k$. As we are interested in implicit variable selection (i.e. shrinking coefficients of unpromising explanatory variables to zero), we implement a modified version of the normal gamma prior, a global local shrinkage prior introduced in \citet{griffin2010inference}:

\begin{equation}
    \beta_{k,p} \sim N(0, \frac{2}{\lambda_k}\tau^2_{k,p})
\end{equation}

where $\tau^2_{k,p}$ denotes the \textit{local} shrinkage parameter of coefficient $p$ in regression $k$. As opposed to \citet{griffin2010inference}, who apply this prior to a standard regression model, we have to deal with $K-1$ separate sets of coefficients in the multinomial logit framework. Thus, we do not use a single global shrinkage parameter $\lambda$, but introduce global shrinkage parameters $\lambda_k$ per equation. This allows for more flexibility and allows for conducting variable selection for each group of the multinomial logit separately. This might be sensible, taking into consideration that the relevant variables responsible for accurately describing class membership might well alter between classes.  Similar prior structures have been implemented into Bayesian time series analysis (see for example \citealp{huber2017adaptive}; \citealp{bitto2016achieving} or \citealp{kastner2016sparse}) and high dimensional spatial models (\citealp{pfarrhofer2019}) recently. To complete the prior setup, we specify the hierarchical structure for $\lambda_k$ and $\tau^2_{k,p}$ to be

\begin{equation}
\begin{split}
\tau^2_{k,p} &\sim G(\theta, \theta) \\
\lambda_k &\sim G(c_0,c_1).
\end{split}
\end{equation}

The priors on the component parameters $\Xi_k$ are application specific. The choice of values for the hyperparameters $c_0$, $c_1$ and $\theta$ is discussed in appendix \ref{hyperparams}.

\subsection{Posterior Simulation}

We implement a Gibbs Sampler to sample the parameters from the full conditional posterior distributions using Markov Chain Monte Carlo (MCMC) methods (\citealp{robert2013monte}). The posterior of the latent class membership indicator $S_i$ is drawn from a multinomial distribution $M(1; p_{i,1},\ldots,p_{i,K})$ with success probabilities $(p_{i,1},\ldots,p_{i,K})$ where

\begin{equation}
\label{eq:classmembershipposterior}
    p_{i,k} = \eta_k(x_i ~|~ \beta)~f_k(y_i ~|~ \Xi_k) ~/~ \sum_{k=1}^K \eta_k(x_i ~|~ \beta)~f_k(y_i ~|~ \Xi_k)
\end{equation}

where $f_k$ denotes the probability density function of the components of the mixture distribution. A posteriori, the regression coefficients are normally distributed with

\begin{equation}
    \pi(\beta_k~|~\cdot) \sim N(m_k, V_k).
\end{equation}

The parameters $m_k$ and $V_k$ of this normal distribution can be derived using the following identities:

\begin{equation}
\begin{split}
        C_{i,k} &= \textrm{log}\sum_{j\neq k}e^{x_i'\beta_j}\\
        C_{k} &= (C_{1,k},\ldots,C_{N,k})\\
        \kappa_{i,k} &= \mathds{1}(S_i = k) - 0.5\\
        \kappa_{k} &= (\kappa_{1,k},\ldots,\kappa_{N,k})\\
\end{split}
\end{equation}

where $\mathds{1}(\cdot)$ denotes the indicator function.\\ 
Let $\omega_{i,k}$ be a latent auxiliary variable that is conditionally P\'{o}lya-Gamma\footnote{For further details on the P\'{o}lya-Gamma sampler as well as standard hyperparameter values and the multinomial setup used here, see \citet{polson2013bayesian} and their technical supplement.} distributed with

\begin{equation}
    \begin{split}
        \omega_{i,k}~|~\beta_j &\sim PG(\phi_i, \psi_{i,k})\\
        \psi_{i,k}& = x_i'\beta_k - C_{i,k}.
    \end{split}
\end{equation}

Using this auxiliary variable, the posterior parameters for sampling $\beta_k$ can be derived\footnote{For details on the derivation see the technical supplement of \citet{polson2013bayesian} as well as for instance \citet[Ch. 3]{koop2003bayesian}.} as

\begin{equation}
    \begin{split}
        V_k^{-1} &= X'\Omega_kX + \textrm{diag}(1/\tau^2_k)\\
        m_k &= V_k(X'(\kappa_k-\Omega_k C_k))\\       
     \end{split}
\end{equation}

where $\Omega_k = \textrm{diag}(\omega_{i,k})$.\\ 
Finally, the posterior distributions of $\lambda_k$ and $\tau^2_{k,p}$ are both of well-known form and can be derived as

\begin{equation}
    \begin{split}
        \pi(\lambda_k~|~\cdot) &\sim G(g_1,d_k)\\
        \pi(\tau^2_{k,p}~|~\cdot) &\sim GIG(\theta-0.5,~ \beta_{k,p}^2,~ \lambda_k \theta)\\
        g_1 &= P * \theta + c_0\\
        d_k &= c_1 + \frac{\theta}{2\sum_{p=1}^P \tau^2_{k,p}},
    \end{split}
\end{equation}

where $P$ is the number of covariates entering the model and $GIG$ denotes the Generalized Inverse Gaussian distribution. \citet{hormann2014generating} provide an efficient adaptive rejection sampling algorithm that makes it possible to easily draw from the $GIG$. This algorithm is implemented in the R package \textit{GIGrvg} (\citealp{leydold2015gigrvg}) which we use in our computations. This completes the simulation setup.

\subsection{Model Selection}
\label{modelselection}
Selecting the number of mixture components still remains a challenging issue. Proposed solutions are the use of reversible jump MCMC algorithms (\citealp{green1995reversible}) or shrinkage on the component weights (\citealp{malsiner2016model}). A further commonly used approach is to estimate the marginal likelihoods of models with different number of components and use these likelihoods to decide how many components are suitable (\citealp{modelselection}).\\
Estimating the marginal likelihood is a non-trivial integration problem that involves a number of possible numerical and computational issues. Starting with the purely statistical problem, we need to compute the marginal likelihood given by 

\begin{equation}
    p(y|M_G) = \int_{\Theta_G} p(y|M_G, \Theta)p(\Theta|M_G)d\Theta_G
\end{equation}

where $M_G$ denotes the model with $G$ components\footnote{The notation differentiates between $G$ and $K$ in this subsection to make it clear that $K$ refers to the number of clusters in the data generating process.} and $\Theta_G = (\Xi_1,\ldots,\Xi_G,\beta_1,\ldots,\beta_{G-1})$ denotes the set of all unknown model parameters in a model with $G$ components\footnote{Note that we assume $\beta_G=0$ to achieve identification in the multinomial logistic framework. Thus, only $G-1$ $\beta$ parameters have to be estimated.}. In the overwhelming majority of cases, this integral does not have a closed form solution. However, several methods may be employed to estimate the value of this integral. We use random permutation bridge sampling to estimate the marginal likelihood for model selection purposes. Bridge sampling was first introduced by \citet{meng1996simulating} and has been thoroughly described for Markov switching and mixture models by \citet{fruhwirth2004estimating}, who concludes that the bridge sampling estimator is the preferable estimator for the marginal likelihood of this model class and superior to related approaches like importance sampling (\citealp{geweke1989bayesian}) or the harmonic mean estimator (\citealp{newton1994approximate}).\\
To get an estimate of the marginal likelihood, we first need to construct an importance density $q(\Theta_G)$ and generate $L$ i.i.d. draws from this density, denoted by $\Theta^{(l)}_G$ with $l = 1,\ldots,L$. This importance density should have the same domain as the posterior distribution and closely resemble the posterior distribution (\citealp{GRONAU201780}). As shown by \citet[Ch. 5.4]{fruhwirth2006}, the bridge sampling estimator can then be derived as 

\begin{equation}
    \hat{p}(y|M_G) = \frac{L^{-1}\sum_{l=1}^L \alpha(\Theta^{(l)}_G) p^{\star}(\Theta_G^{(l)}|y,M_G)}{M^{-1}\sum_{m=1}^M \alpha(\Theta^{(m)}_G)q(\Theta^{(m)}_G)}
\end{equation}

where $\Theta^{(m)}_G$ with $m=1,\ldots,M$ are the $M$ posterior draws from the Gibbs sampler output using $G$ components and $p^{\star}(\cdot)$ denotes the non-normalized posterior distribution. The choice of $\alpha(\Theta_G)$ is arbitrary, however, \citet{meng1996simulating} discuss an asymptotically optimal choice which minimizes the expected relative error of the estimator. It is given by

\begin{equation}
    \alpha(\Theta_G) \propto \frac{1}{Lq(\Theta_G) + Mp(\Theta_G|y,M_G)}.
\end{equation}

The bridge sampling estimate of the marginal likelihood $\hat{p}_{BS}$ can be obtained using the following algorithm:

   \begin{enumerate}
       \item Run the MCMC sampler and save $M$ posterior draws $\Theta_G^{(m)}$ from the mixture posterior $p(\Theta_G|y,M_G)$ where $m=1,\ldots,M$.
       \item Construct an importance density $q(\Theta_G)$ and generate $L$ i.i.d. samples $\Theta_G^{(l)}$ from the importance density.
       \item Choose a starting value for $\hat{p}_{BS,0}$.
       \item Run the following recursive process until convergence is achieved:
   \end{enumerate}
  \begin{equation}
           \hat{p}_{BS,t+1} = \frac{L^{-1}\sum_{l=1}^L \frac{p^{\star}(\Theta_G^{(l)}|y,M_G)}{Lq(\Theta_G^{(l)}) + Mp^{\star}(\Theta_G^{(l)}|y,M_G) / \hat{p}_{BS,t}}
 }{M^{-1}\sum_{m=1}^M  \frac{q(\Theta_G^{(m)})}{Lq(\Theta_G^{(m)}) + Mp^{\star}(\Theta_G^{(m)}|y,M_G) / \hat{p}_{BS,t}}}
  \end{equation}
       
In general, both the construction and the evaluation of the importance density for mixture-of-experts models are challenging due to the multimodal nature of the likelihood function. We follow the approach described in \citet{fruhwirth2004estimating}, who states that the importance density for mixture models can be constructed in a fully automatic manner by saving the posterior distribution moments of $S$ randomly selected MCMC draws. In a random permutation sampler, this results in a multimodal importance density that approximates the modes of the posterior distribution. An i.i.d. sample from this importance density can then be generated by drawing from a uniform mixture of the $S$ saved densities. Additional details on the construction of an importance density and the implementation of a bridge sampler for the proposed model are provided in appendices \ref{bridge} and \ref{numeric}.\\
To compute a marginal likelihood estimate using this procedure, it is necessary to choose a starting value for the bridge sampler. Reasonable choices include alternative estimates of the marginal likelihood. \citet[Ch. 5.4.6]{fruhwirth2006} suggests using the importance sampling estimator or the reciprocal importance sampling estimator  of the marginal likelihood. Both estimators can be derived from the same functional values that are needed to compute the bridge sampling estimate\footnote{However, other starting values are possible. \citet{GRONAU201780} choose $0$ as starting value, stating that ''usually the exact choice of the initial value does not seem to influence the convergence of the bridge sampler much.''.}.\\
Note that in order to evaluate the non-normalized posterior distribution, it is necessary to use the marginal prior densities of the parameters that are specified using a hierarchical prior setup. The marginal prior of $\beta_{k,p}$ is available in closed form and can be derived as

\begin{equation}
\small
\label{eq:proof}
  p(\beta_{k,p}) = \frac{\sqrt{\theta\lambda^2}^{~\theta+0.5}}{\sqrt{\pi}~2^{\theta-0.5}~\Gamma(\theta)}|\beta_{k,p}|^{\theta-0.5}K_{\theta-0.5}(\sqrt{\theta\lambda^2}~~|\beta_{k,p}|),
\end{equation}

where $K_x(\cdot)$ is the modified Bessel function of the second kind with index $x$ and $\Gamma(\cdot)$ is the gamma function (see for instance \citealp{bitto2016achieving}).\\
A thorough discussion of the bridge sampling technique is out of scope of this article. However, so far literature has been rather sparse on the practical computation of bridge sampling estimates in the context of mixture models and especially mixture-of-experts models. An exception is the recent review by \citet[Section 12.3.3]{gor-fru:mod} who give details on the procedure for mixture-of-experts models. 

\subsection{Label Switching and Identification}
\label{ls}
Parameter estimation in this model family poses various difficulties, especially in a Bayesian framework. Label switching is a known issue when estimating mixture models (\citealp{hurn2003estimating}; \citealp{jasra2005markov}). It is the result of the multimodal likelihood function being invariant to relabeling the components as pointed out by \citet{redner1984mixture}. This can be problematic as switching labels during MCMC sampling might result in heavily distorted, multimodal posterior distributions that are difficult to summarize. Deriving point estimates such as posterior means then becomes inappropriate (\citealp{stephens2000dealing}).\\ 
Early approaches deal with label switching by introducing simple restrictions on the mixture parameters such as $\eta_1 < \ldots < \eta_K$ (see for instance \citealp{lenk2000bayesian}). However, identifying simple restrictions in high-dimensional models might be cumbersome or infeasible. In addition, if the restriction does not result in the MCMC sampler visiting all modes of the multimodal likelihood evenly, estimates of the marginal likelihood of the model might be biased according to \citet{fruhwirth2004estimating}.\\
Early references for other relabeling algorithms include \citet{celeux1996stochastic}. However, their suggestions require known true parameter values, which makes them difficult to apply in real data settings. \citet{stephens2000bayesian} suggests to relabel the draws such that the marginal parameter posterior distributions are as unimodal as possible. \citet{stephens2000dealing} provides a literature review as well as a decision theoretic framework to deal with label switching.\\
We employ the approach described in \citet[Ch. 3.7.7]{fruhwirth2006} and identify the posterior draws using a postprocessing procedure via $k$-means clustering. In addition, to force the sampler to explore the full mixture posterior distribution, random permutation sampling is introduced (\citealp{fruhwirth2001markov}). That is, every MCMC iteration is concluded by a random permutation step before storing the parameter draws to achieve balanced label switching.\\
The identification algorithm employed is based on the idea of clustering the parameter draws using distance based measures in the point process representation of the MCMC output. After $M$ saved unconstrained MCMC iterations, $k$-means clustering is applied to all $MK$ posterior draws within a suitable parameter subset. The idea is that draws belonging to the same mixture component will be sorted into the same group by the clustering algorithm. The permutation sequence that results from this $k$-means procedure can then be used to reorder the posterior draws and obtain unique identification for further parameter inference. More formally, we use the following two block algorithm:

   \begin{enumerate}
      \item MCMC Sampling
      \begin{enumerate}
         \item Simulate parameters $\Theta^{(t)}$ conditional on the classification sequence $S^{(t-1)}$.
         \item Classify each observation $y_i$ conditional on $\Theta^{(t)}$.
         
         \item Select one of the $K!$ possible permutations of the component labels randomly. Use the resulting labeling sequence $\rho^{t}(1),\ldots,\rho^{t}(K)$ to relabel both the parameter draw $\Theta^{(t)}$ and the classification sequence $S^{(t)}$.
      \end{enumerate}
      \item Identification
      \begin{enumerate}
          \item Arrange the MCMC draws in a matrix with $MK$ rows and $r$ columns, where $r$ denotes the number of parameters deemed necessary to identify the model after for instance visually inspecting the MCMC output.\footnote{It can be shown that identifying a mixture model using a mere subset of the parameter space fully identifies the model.}
          \item Cluster all $MK$ draws using $k$-means centroid analysis. 
          \item For each MCMC draw $m=1,\ldots,M$, construct a classification sequence $\rho^{(t)}$ of size $K$ containing information on cluster membership for each parameter draw.
          \item Check whether $\rho^{(t)}$ is one of the $K!$ possible permutations of $(1,\ldots,K)$. If this is not the case, remove the draw.
          \item All remaining draws can be identified through reordering using the classification sequences $\rho^{(t)}$, which guarantees unique labeling. Consequently, the identified draws can be used for parameter inference.
      \end{enumerate}
   \end{enumerate}

Step 2(d) is implemented to ensure that we only use draws where a unique labeling can be found. By removing draws where $\rho^{(t)}$ is not a permutation of $(1,\ldots,K)$, we remove draws where clusters are overlapping in the point process representation. When two or more clusters overlap, no unique labeling for each of the $K$ parameter draws in MCMC draw $m$ is achievable through $k$-means centroid analysis. The ratio of removed draws to the number of saved MCMC draws can be used as an indicator for how well the model is able to separate the mixture clusters. A high rate of non-permutations usually points in the direction of an over-fitting model. For further information on this identification algorithm, refer for instance to \citet[Appendix 2]{malsiner2016model}. For further and more specific information on the identifiability of mixture-of-experts models, see for instance \citet{jiang1999identifiability} or the excellent discussion with many examples in \cite{gor-fru:mod}.

\section{Simulation Study}

To illustrate the performance of the proposed prior structure, we conduct a number of simulation studies to compare our approach to other possible model setups. The normal gamma shrinkage prior is compared to the standard prior setup suggested in \citet{polson2013bayesian} and the stochastic search variable selection prior (SSVS; \citealp{george1993variable}). The basic concept of the SSVS prior is similar to ours in terms of model structure and computational approach. Therefore, a simulation based comparison of the two models seems advisable. The SSVS prior relies on the idea of specifying a mixture of two normal densities as prior for each multinomial logit coefficient. Both normal densities are centered at 0. One has a large variance (''slab'') while the other one has a small variance (''spike''). Using standard mixture modeling techniques, it is possible to estimate whether a particular coefficient will be drawn from the slab or from the spike component of the mixture. Formally, we specify 

\begin{equation}
    \beta_{k,p} \sim (1-\delta_{k,p}) N(0,\zeta_1^2) + \delta_{k,p} N(0,\zeta_2^2)
\end{equation}

where $\zeta_1^2 << \zeta_2^2$ and $\delta_{k,p}$ is the binary inclusion indicator of covariate $p$ in group $k$. For details, see \citet{george1993variable}. Following \citet{ghosh2011}, we set $\gamma_2^2 = 1$. The normal spike component is specified with variance $\gamma_1^2 = 0.01$.\\
A variety of simulation exercises is conducted. The first study evaluates the performance of the prior only. That is, the relative performance of the NG prior is explored in a multinomial logistic regression setup. In a second step, the three priors are compared in various classification problems where they are employed to cluster observations arising from bivariate normal distributions. Overall, the simulation studies suggest a rather similar performance of the SSVS prior and the NG prior when it comes to estimating the coefficients in the class membership part of the model. However, the NG prior usually shows slight benefits, especially in shrinking unnecessary coefficients to zero, in high sparsity settings and when estimating marginal likelihoods. Details are discussed below.

\subsection{Prior Performance}
\label{prior_study}
In this subsection, the prior of \citet{polson2013bayesian} applying no shrinkage (hereafter ''Standard Prior''\footnote{We set the prior variance of the standard prior to 10 as proposed in \citet{polson2013bayesian}.}) and the SSVS prior are compared to the NG prior in a multinomial logistic regression setup. This preliminary analysis allows us to evaluate the shrinkage performance independently of the mixture setup.\\
Using the data generating process in Eq.~\hspace{-1mm}\eqref{eq:mnlogit}, we simulate four groups with 750 observations and 20 explanatory variables each. The true parameter vectors are chosen to be sparse, thus creating the need for considerable shrinkage within the estimation of the multinomial logistic regression. The true coefficient values are $\beta_1 = (0.8, 1, 2, 0.5, 0, \ldots)'$, $\beta_2 = (0.3, 0, 0, 0, -1, 1.7, -2, 0, \ldots)'$ and  $\beta_3 = (0.3,1,-2,0.8,0.9, 0, \ldots)'$\footnote{This setup corresponds to the simulation study conducted in \citet{ghosh2011}}. All explanatory variables are drawn from a standard normal distribution. Note that this setup implies that we need to deal with group specific relevant membership predictors. Thus, to obtain good estimates, group specific shrinkage is necessary. As this simulation uses a large number of observations, a quite informative likelihood results. Hence, we extend the setup described above by two scenarios using 300 and 100 observations, respectively. This should enable us to evaluate the prior performance in an environment with comparatively uninformative data. We implement a Gibbs sampler using 25000 draws after a burn-in period of 5000 draws. The mean estimates of 25 simulation runs are then compared.\\

\begin{table*}[!h]
\centering
\begin{threeparttable}
        \caption{Simulation Study I Results}

\label{tab:simmain}

 \footnotesize
        \setlength\tabcolsep{3pt}
        
        \renewcommand\arraystretch{0.5}

  \begin{tabular}{ld{3.3}d{3.3}d{3.3}d{3.3}d{3.3}}
    \toprule

          & \thead{RMSE} & \thead{RMSE} & \thead{RMSE} & \thead{RMSE} & \thead{Time} \\
        & \thead{(Zeros)} & \thead{(Non Zeros)} & \thead{(Overall)} & \thead{(P.P.)} & \thead{(in sec.)} \\
\textbf{N = 3000}       &               &                   &                &             &      \\
Standard Prior &        2.85 &1.06& 1.71& 1.57& 0.97 \\ 
SSVS           &        1.82 &1.00 &1.26 &1.28 &0.97 \\
NG             &         1.00&1.00&1.00&1.00&1.00 \\ 
\midrule
\textbf{N = 300}        &               &                   &                &             &      \\
Standard Prior &       3.89 &2.07& 2.63& 1.71& 0.67 \\ 
SSVS           &       1.28 &0.87& 0.98& 1.06& 0.83 \\ 
NG             &       1.00&1.00&1.00&1.00&1.00 \\  
\midrule
\textbf{N = 100}        &               &                   &                &             &      \\
Standard Prior &       7.36 &4.71& 5.52& 1.80& 0.51 \\ 
SSVS           &       1.17 &0.69& 0.84& 1.04& 0.74 \\ 
NG             &       1.00 &1.00& 1.00& 1.00&1.00 \\
%Standard Prior &        6.16 &   8.83 &   6.69 &   3.60 & 448.56 \\ 
%SSVS           &        3.94 &   8.35 &   4.94 &   2.94 & 449.60 \\
%NG             &         2.16 &   8.35 &   3.92 &   2.29 & 461.41 \\ 
%\midrule
%\textbf{N = 300}        &               &                   &                &             &      \\
%Standard Prior &       29.19 & 63.39 & 37.31 & 12.94 & 57.30 \\ 
%SSVS           &       9.62 & 26.55 & 13.94 &  7.99 & 70.80 \\ 
%NG             &       7.51 & 30.57 & 14.19 &  7.55 & 85.50 \\ 
%\midrule
%\textbf{N = 100}        &               &                   &                &             &      \\
%Standard Prior &       119.16 & 286.90 & 159.50 &  23.29 &  28.74 \\ 
%SSVS           &       18.90 & 41.97 & 24.39 & 13.46 & 41.81 \\ 
%NG             &       16.20 & 60.93 & 28.87 & 12.96 & 56.32 \\
\bottomrule
\end{tabular}
\begin{tablenotes}
\item\textit{Note:} The estimates correspond to the average value across 25 runs. RMSEs are reported relative to NG RMSEs.
\end{tablenotes}
\end{threeparttable}
\end{table*}

\renewcommand\linespread{1.5}

\begin{figure}[!t]
\centering
\subfloat[Standard Prior]{\includegraphics[width = 0.33\textwidth]{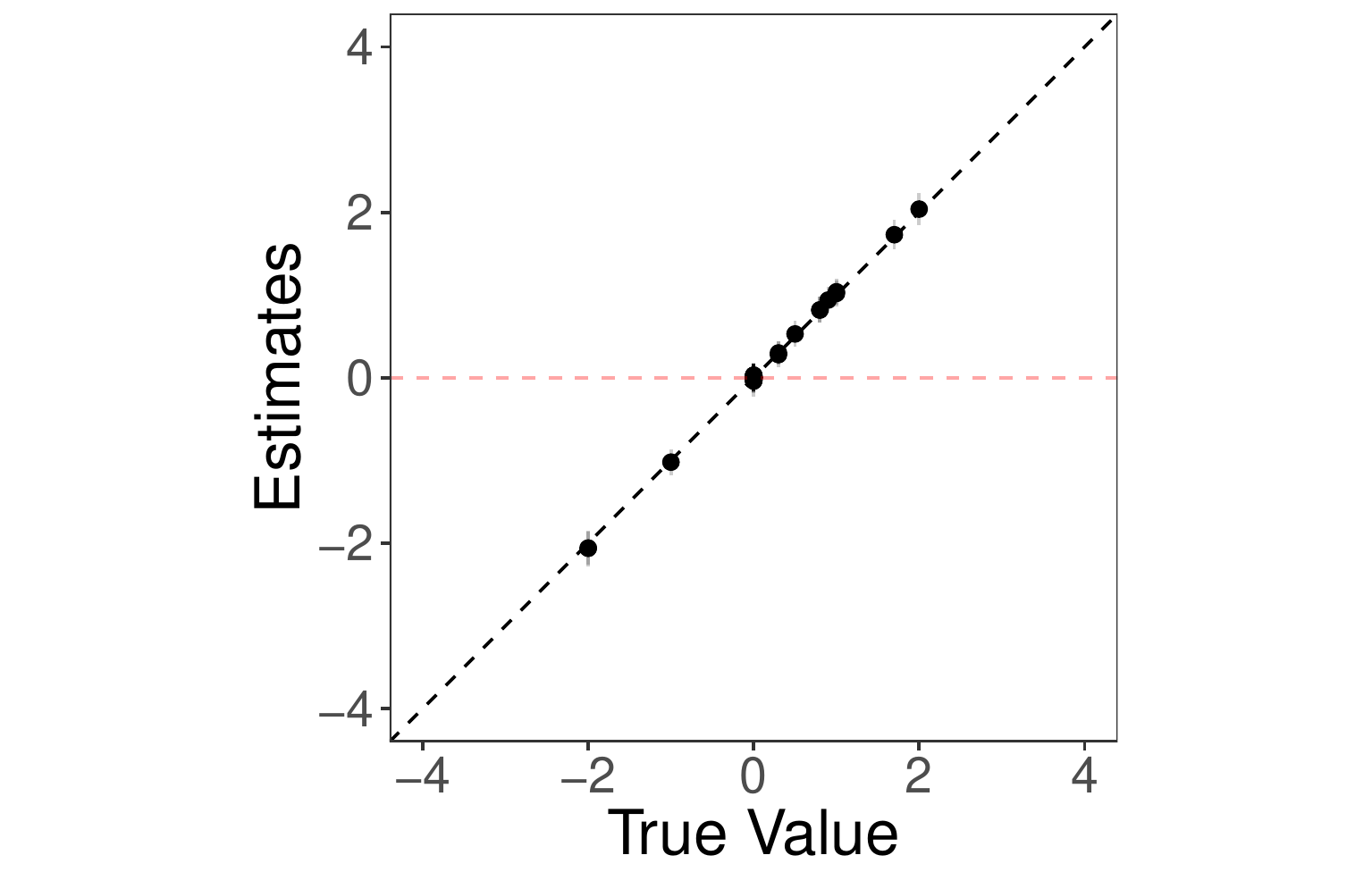}}
\subfloat[SSVS Prior]{\includegraphics[width = 0.33\textwidth]{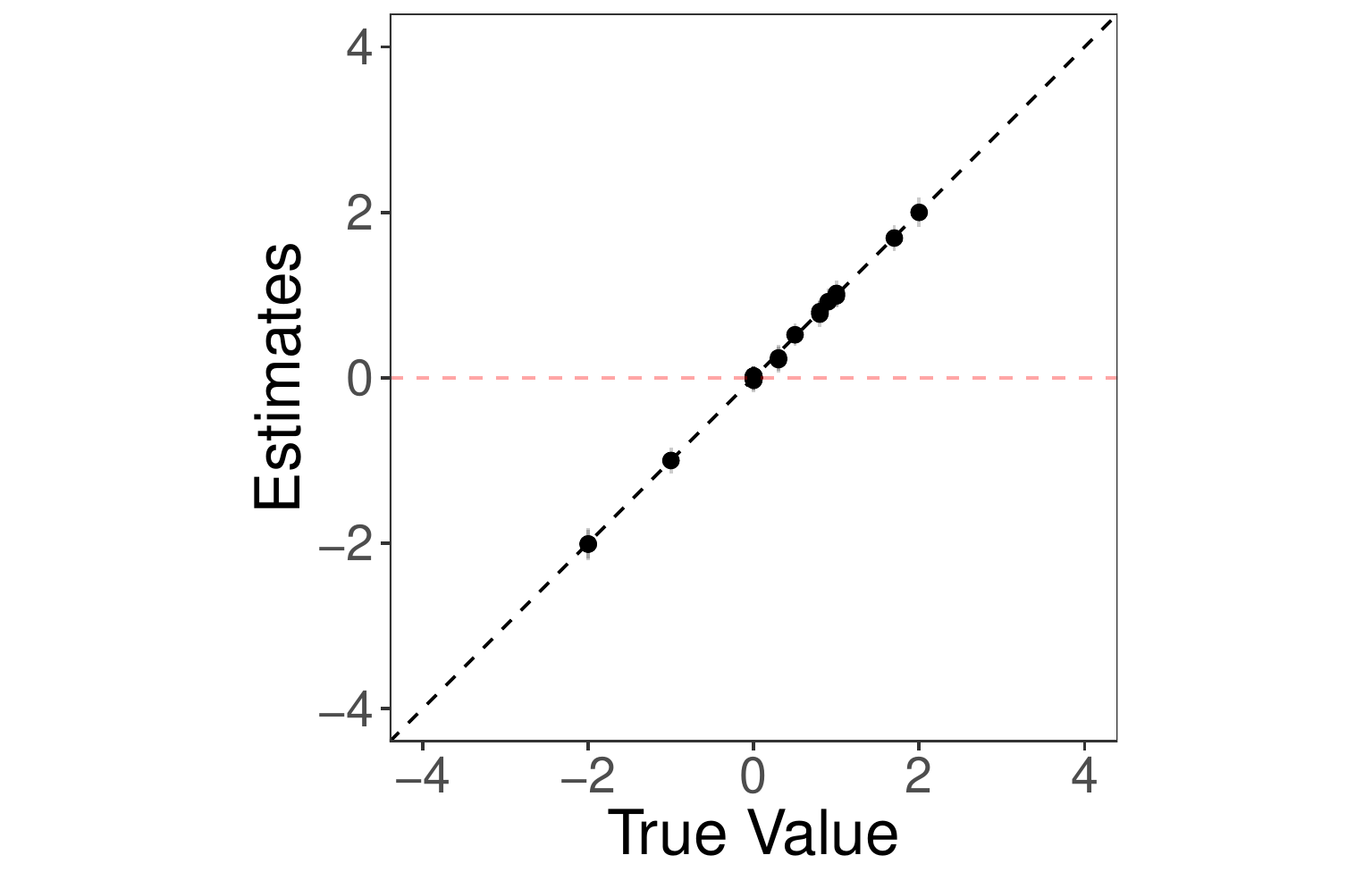}}
\subfloat[NG Prior]{\includegraphics[width = 0.33\textwidth]{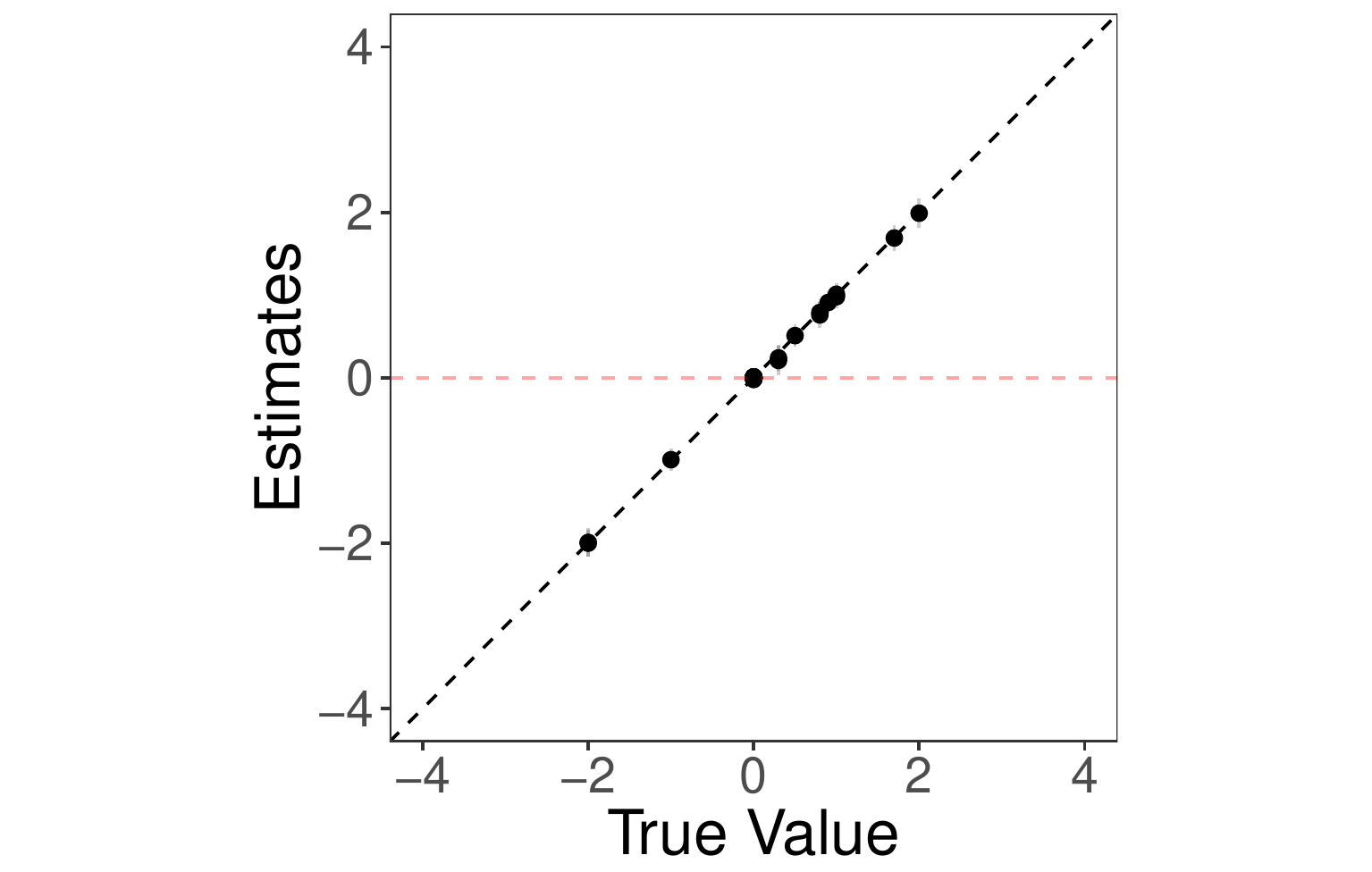}} 
\caption{Posterior mean estimates vs. true values for different model setups (with dashed 45$^{\circ}$ line) for N = 3000}
\label{fig:trueestimates1}
\end{figure}

\begin{figure}[!t]
\centering
\subfloat[Standard Prior]{\includegraphics[width = 0.33\textwidth]{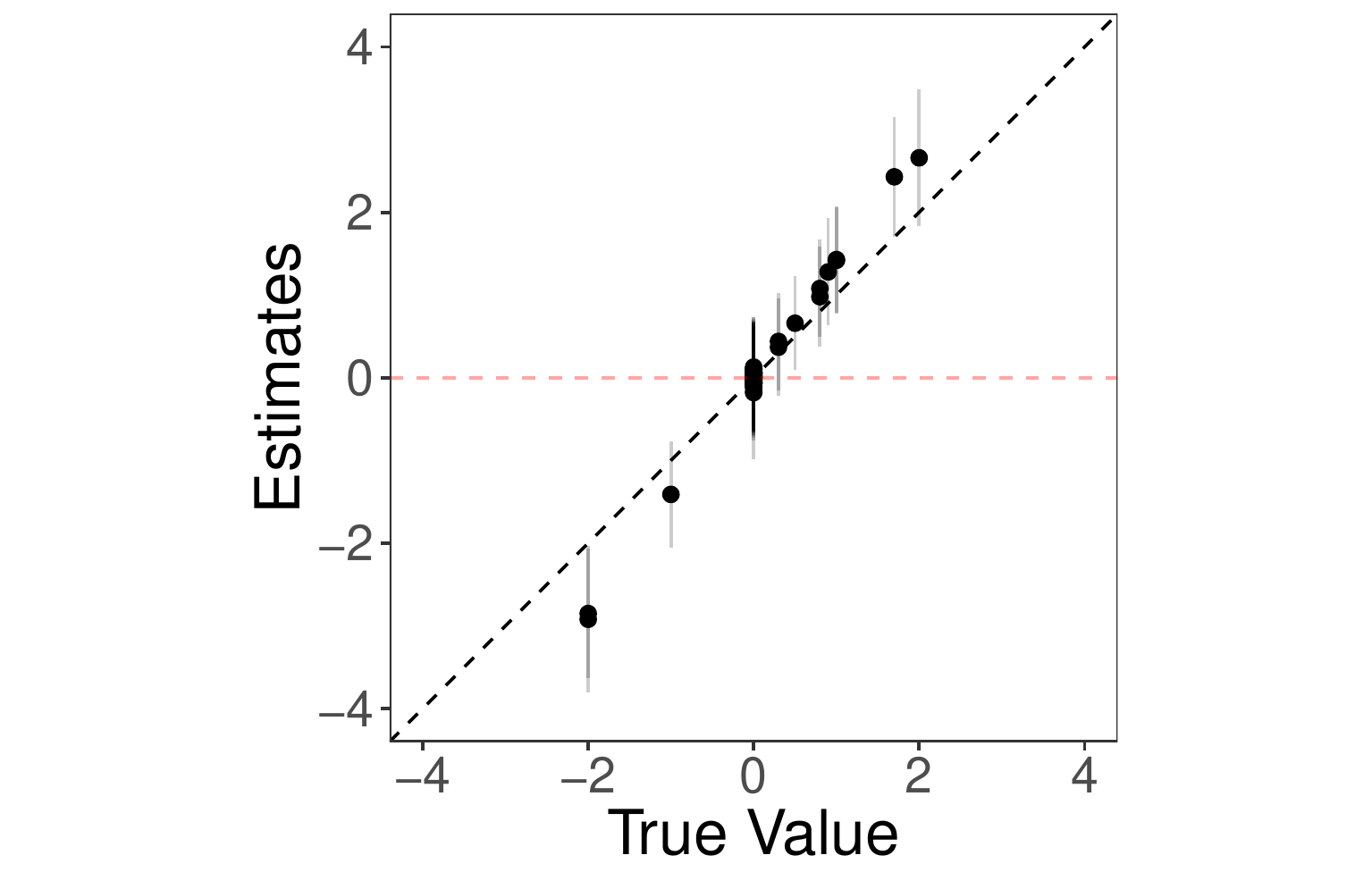}}
\subfloat[SSVS Prior]{\includegraphics[width = 0.33\textwidth]{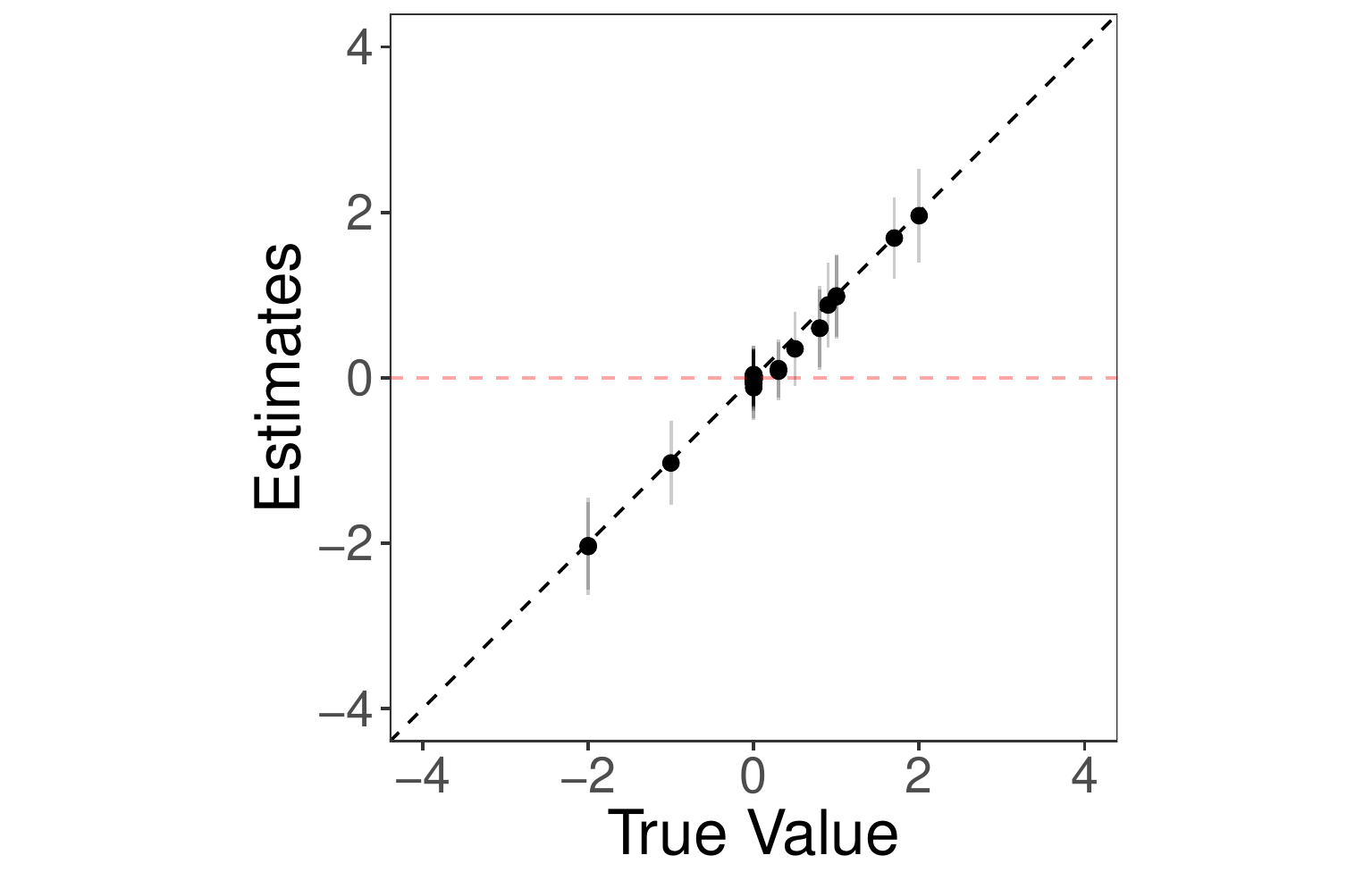}}
\subfloat[NG Prior]{\includegraphics[width = 0.33\textwidth]{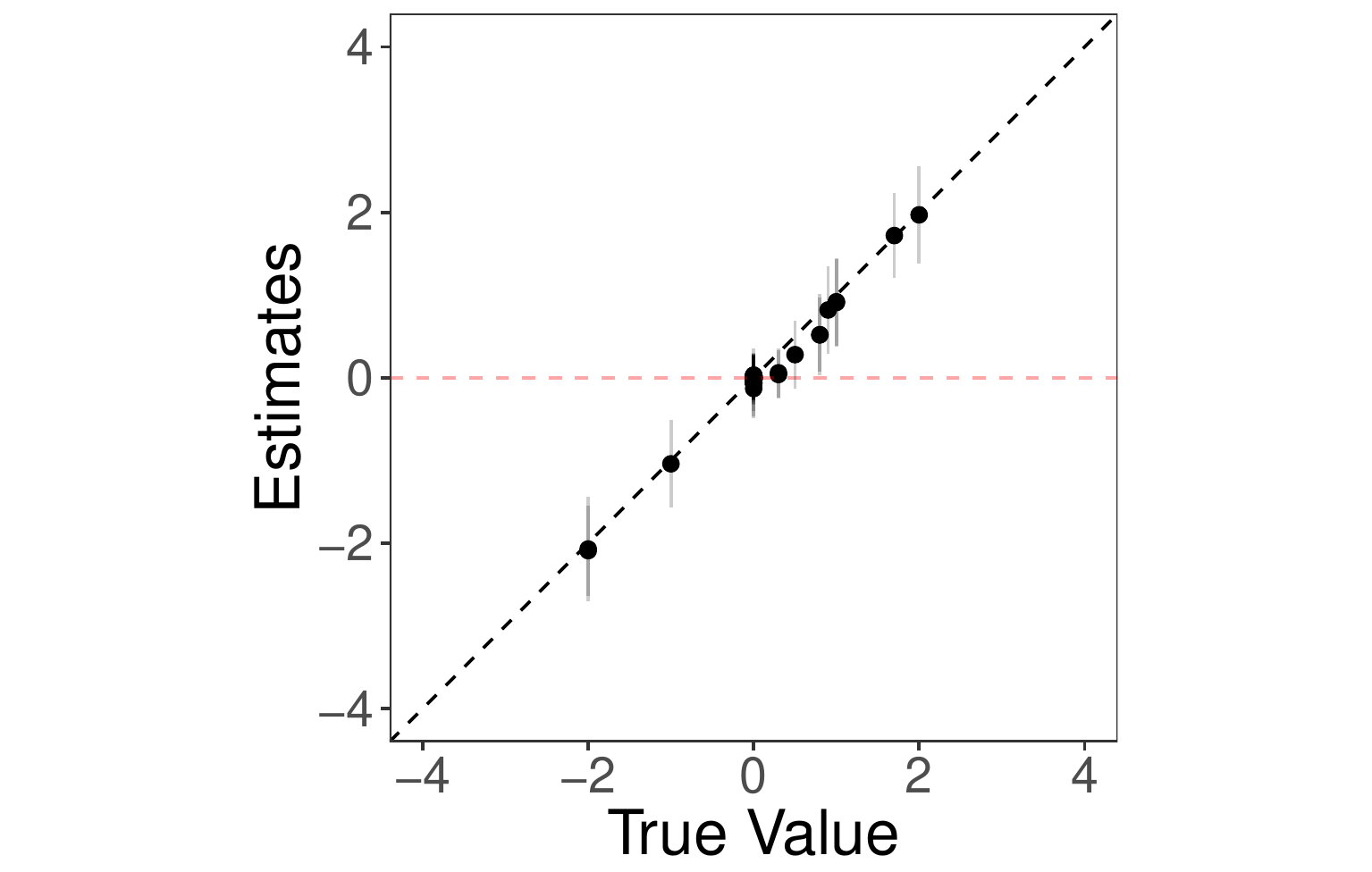}} 
\caption{Posterior mean estimates vs. true values for different model setups (with dashed 45$^{\circ}$ line) for N = 300}
\label{fig:trueestimates2}
\end{figure}

\begin{figure}[!t]
\centering
\subfloat[Standard Prior]{\includegraphics[width = 0.33\textwidth]{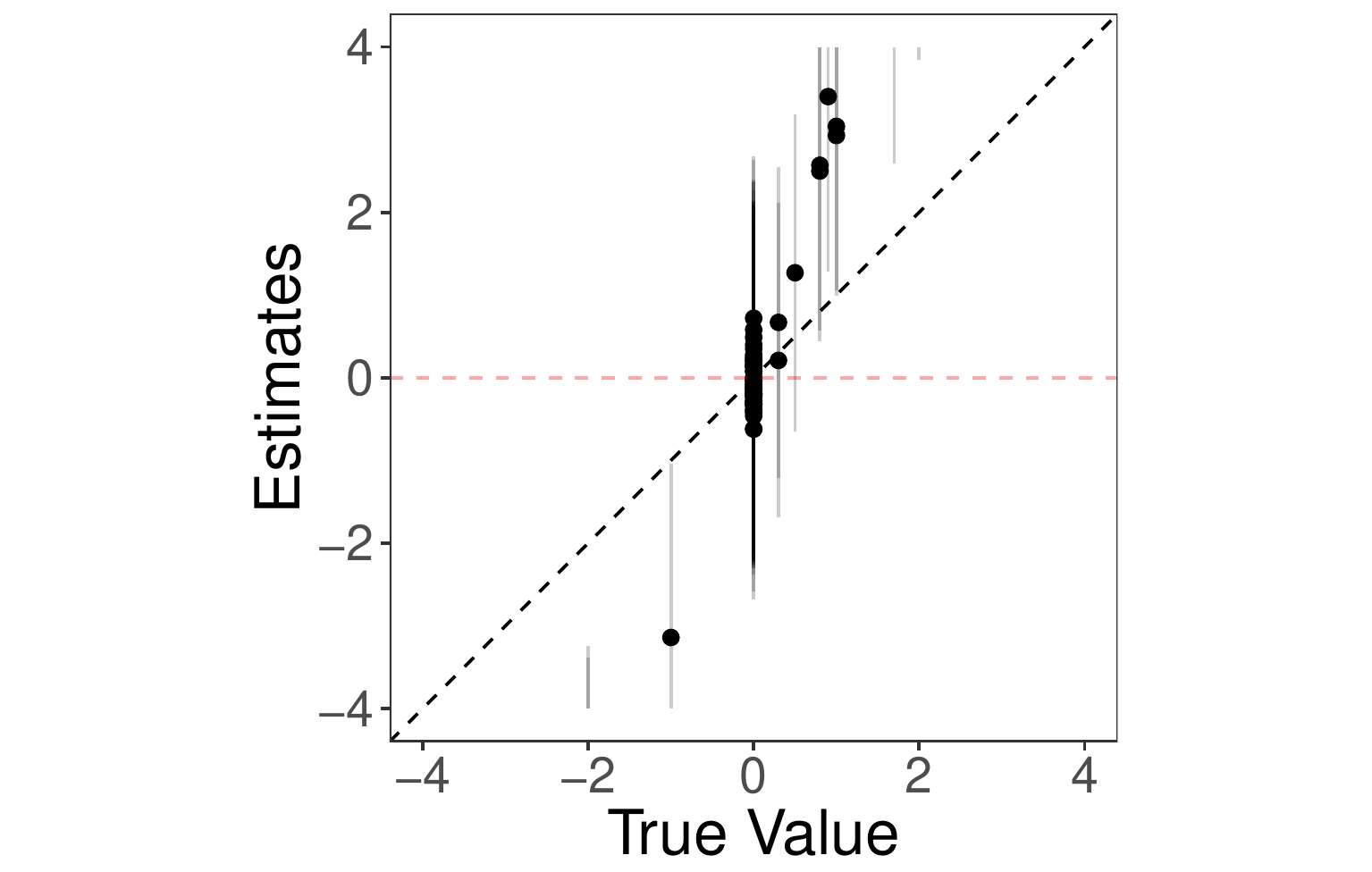}}
\subfloat[SSVS Prior]{\includegraphics[width = 0.33\textwidth]{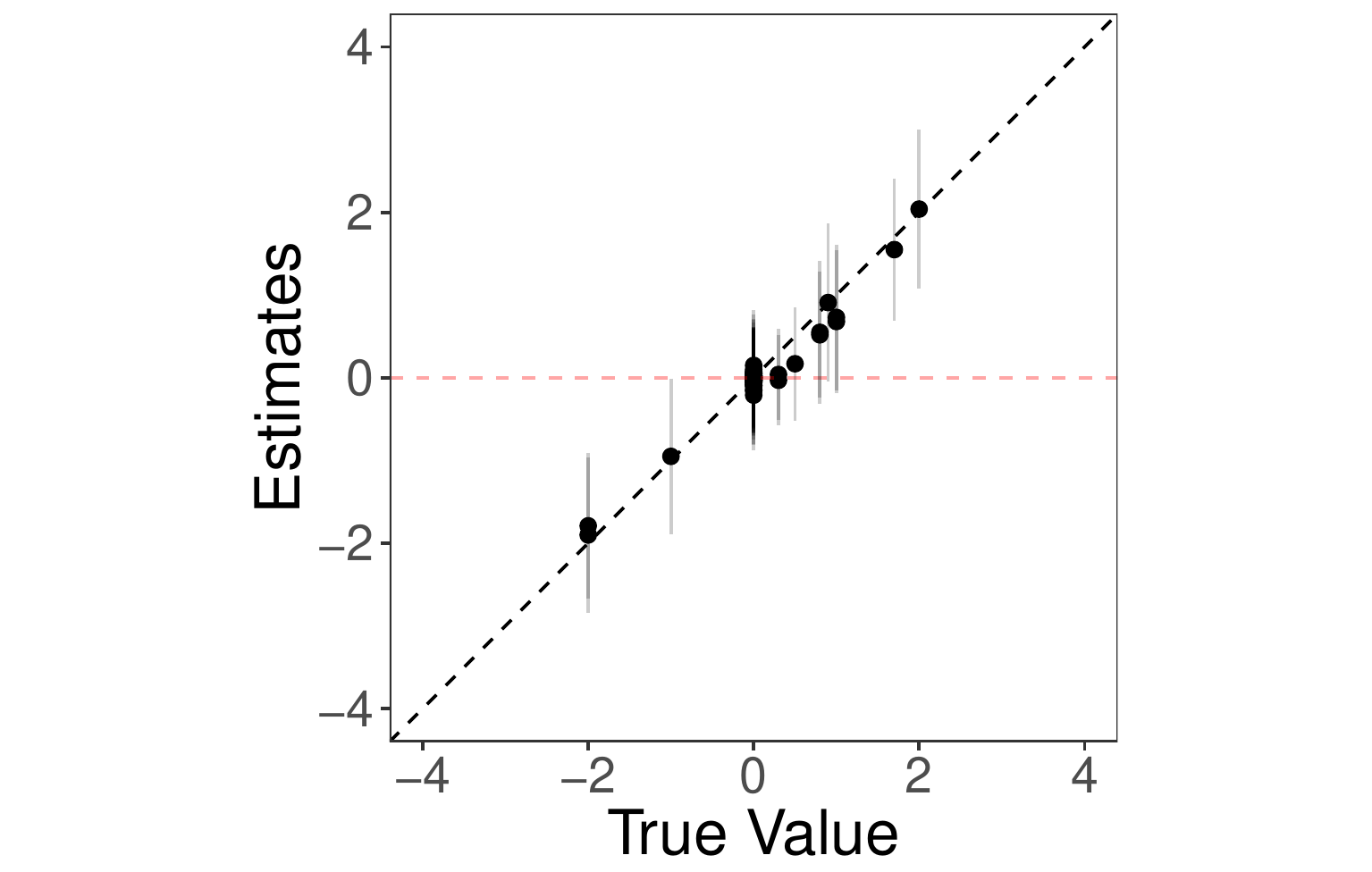}}
\subfloat[NG Prior]{\includegraphics[width = 0.33\textwidth]{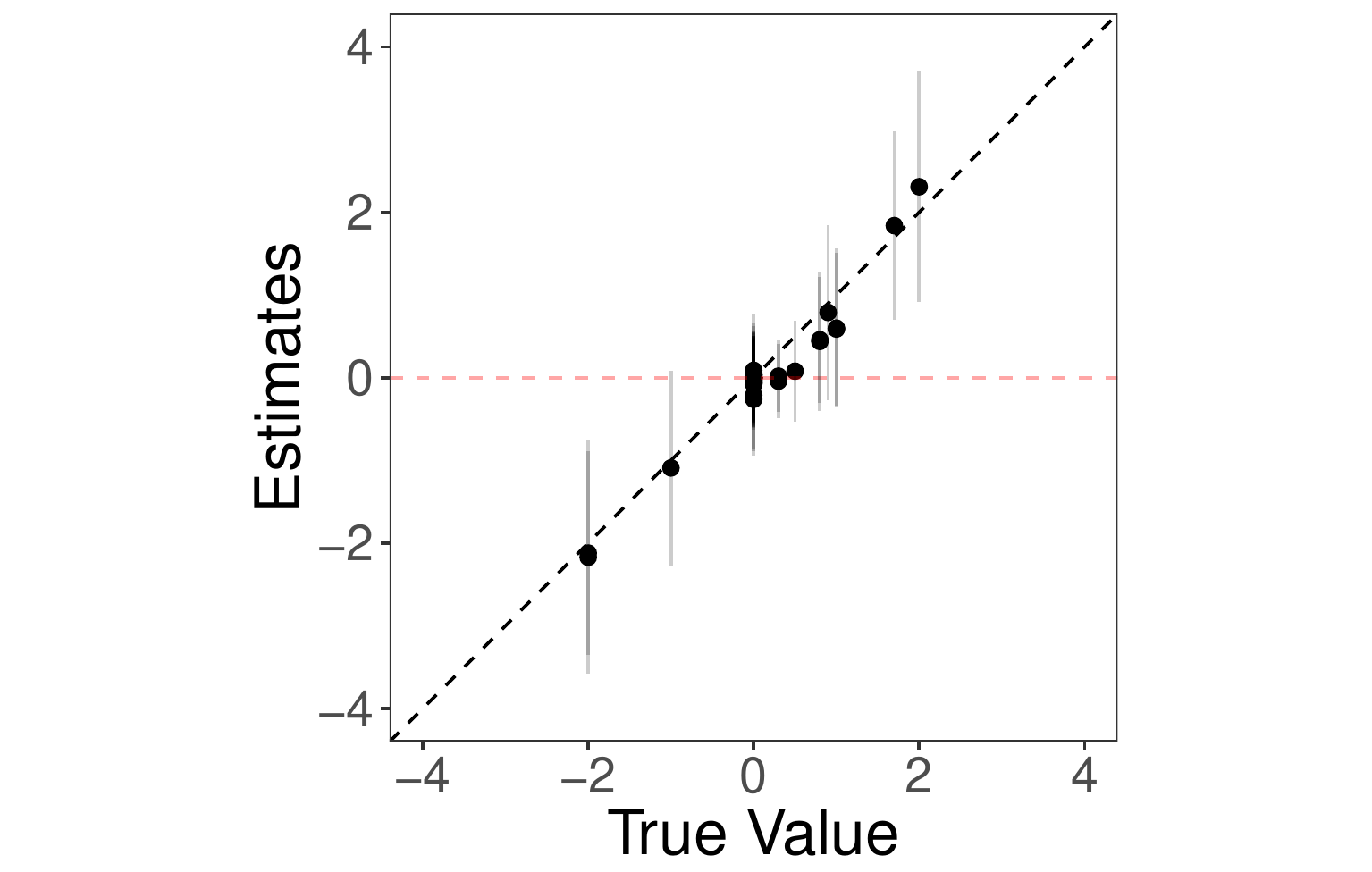}} 
\caption{Posterior mean estimates vs. true values for different model setups (with dashed 45$^{\circ}$ line) for N = 100}
\label{fig:trueestimates3}
\end{figure}

Table \ref{tab:simmain} reports the root mean squared error (RMSE) with respect to the coefficients that are truly zero, the coefficients that are truly different from zero, all coefficients and the predicted probabilities (P.P., defined in Eq. \ref{eq:classmembershipposterior}) resulting from the estimation. This enables us to separately examine how well the priors are able to shrink unimportant coefficients to 0, how precise the point estimates are and whether they are able to give useful estimates of the predicted probabilities. These predicted probabilities are of utter importance in the mixture-of-experts framework, as they will directly influence class membership and therefore all estimated model parameters.\\
The results suggest that the first simulation using 3000 observations is not a very competitive environment. The likelihood is quite informative, resulting in precise estimates even for the standard setup without introducing shrinkage. Figure \ref{fig:trueestimates1} plots the true values against the posterior mean estimates of the respective models. The uncertainty surrounding the posterior means is given by an interval of $\pm 1.96 * SD$ where $SD$ is the posterior standard deviation. Scatterplots suggest that all three models are able to revocer the true coefficient values well. Nevertheless, the NG setup performs particularly well and even outperforms the SSVS setup in terms of precision. However, it comes at the cost of a slightly prolonged computation time.\\
Using just 10\% of the observations, estimation becomes more difficult as the data becomes less informative as seen in Figure \ref{fig:trueestimates2}. The point estimates become considerably worse. The standard prior has problems to recover the true values, as the enlarged RMSEs indicate. The NG prior shows slight advantages in terms of shrinkage and in predicting cluster membership probabilities. However, the SSVS setup is able to provide more accurate point estimates and therefore has a slightly lower RMSE with respect to the true non-zero coefficients and regarding the overall coefficient RMSE.
Further reducing the number of observations to $N=100$ leads to inflated coefficient estimates and increasing uncertainty when applying no shrinkage, as depicted in Figure \ref{fig:trueestimates3}. This results in enlarged RMSEs. The performance of the SSVS and NG prior remains rather similar to the case with $N=300$, however, the shrinkage priors also show larger uncertainty surrounding the posterior means. The SSVS prior produces better point estimates, but is not as efficient as the NG prior when it comes to shrinking unnecessary coefficients to zero. The NG prior shows a slightly better performance when predicting the class membership probabilities. We would like to note that both the NG prior and the SSVS prior perform rather well in absolute terms, producing small RMSEs in general\footnote{It should also be noted that the performance of the analyzed priors depends on the imposed prior variances. However, simulations using different variances did not change the results qualitatively. See \citet{sfswagner} for a thorough discussion and comparison of various shrinkage priors.}. All coefficient estimates of these simulations are provided in Tables 4-6 in Appendix \ref{simstudy}. The performance of the three priors in a full mixture-of-experts classification setting are discussed in the next subsection.\\
\renewcommand\baselinestretch{1}
\\
\renewcommand\baselinestretch{1.5}

\subsection{Classification Exercises}

\begin{figure}[!t]
\centering
\subfloat[Well Separated]{\includegraphics[width = 0.3\textwidth]{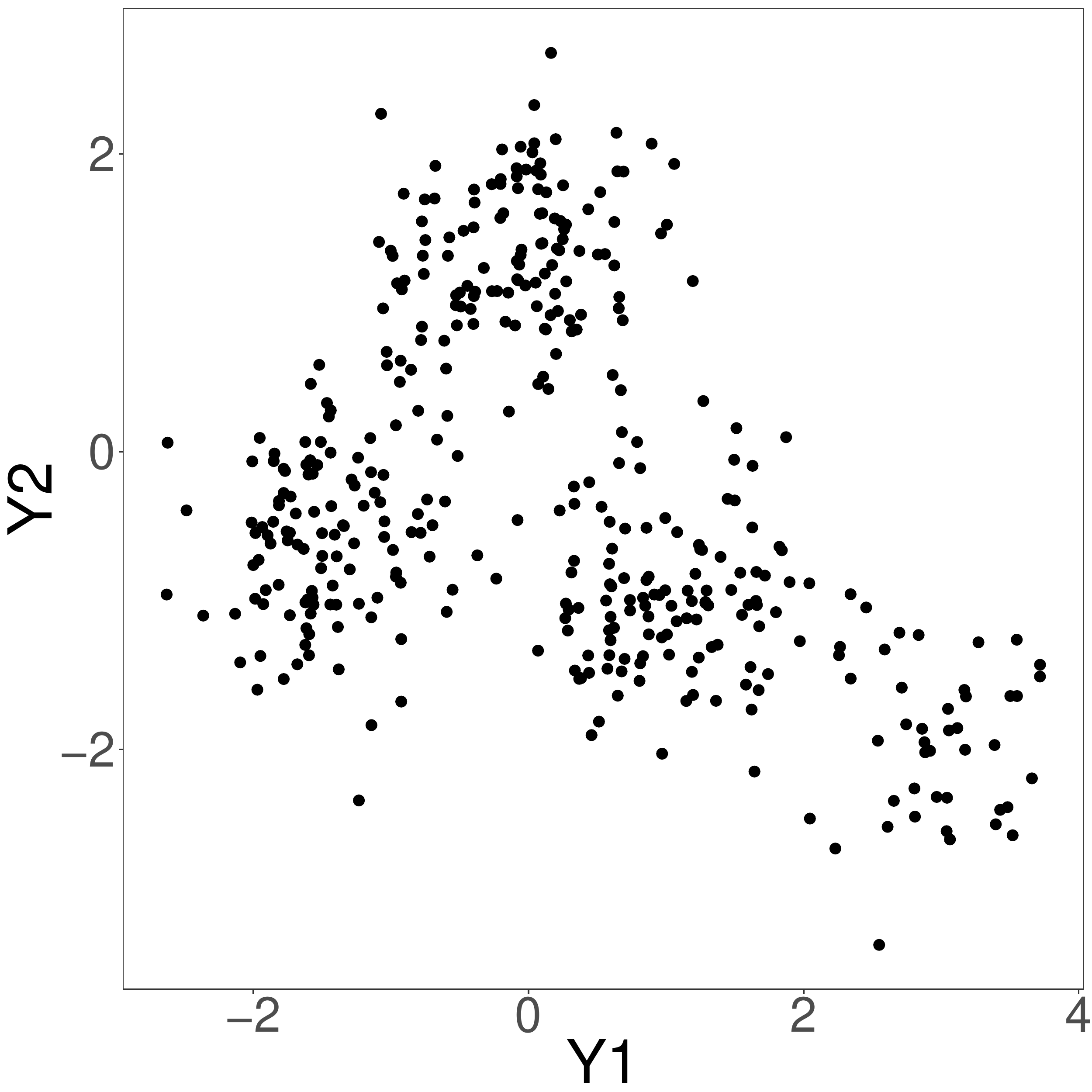}}
\subfloat[Overlapping]{\includegraphics[width = 0.3\textwidth]{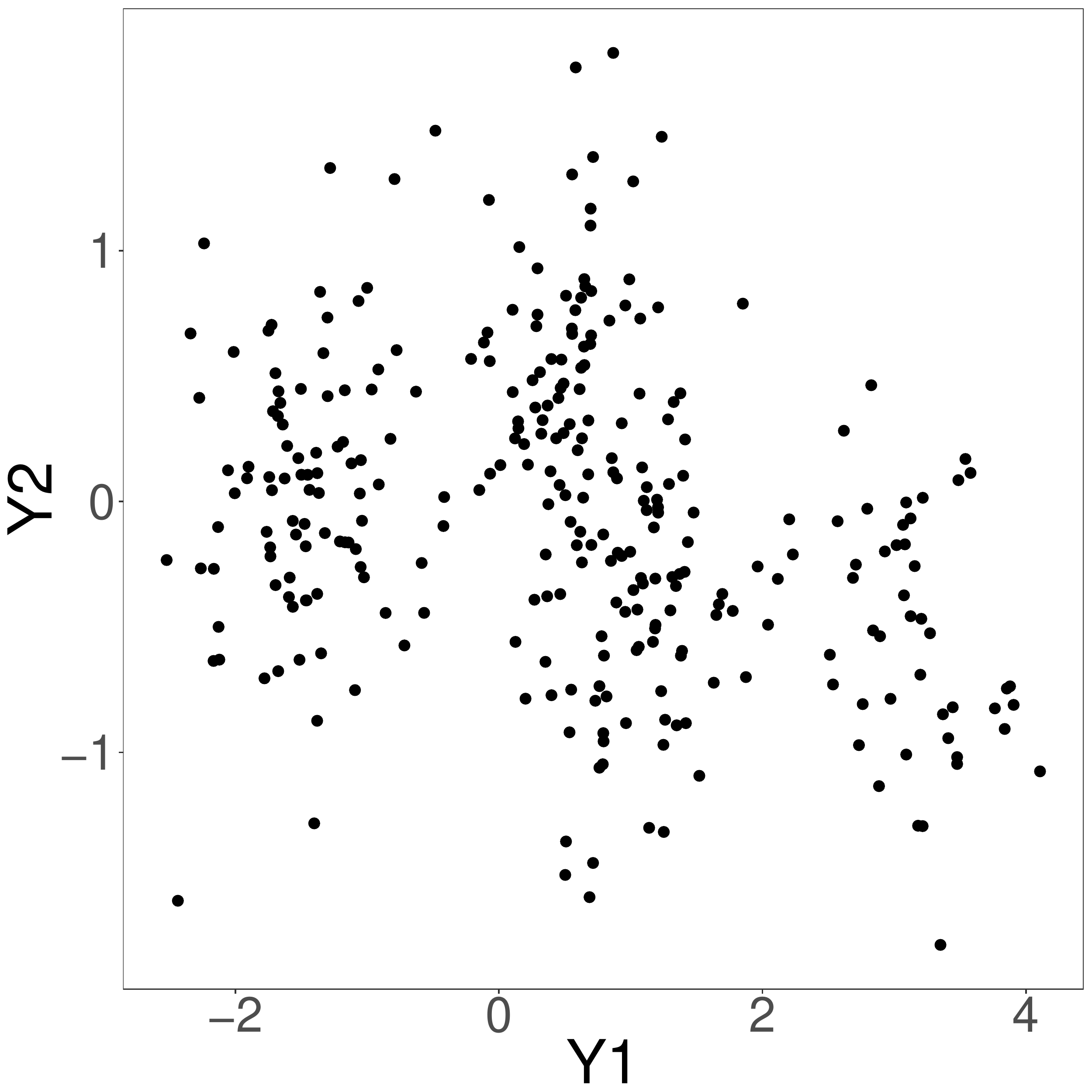}}
\caption{Example datasets for the ''well separated'' and ''overlapping'' simulation scenarios.}
\label{fig:data_simulations}
\end{figure}

To further examine the performance of the priors, four simulation studies in a full mixture-of-experts framework are conducted. These simulations differ from each other with respect to the degree of overlapping of the clusters, the number of regressors and the complexity of the sparsity structure in the true coefficient vectors.\\
Datasets with 300 observations and four clusters arising from bivariate normal distributions are generated. We differentiate between a ''well separated'' case and an ''overlapping'' case using $\mu_{1} = (-1.5,-0.5)'$, $\mu_{2}=(0,1.3)'$, $\mu_{3}=(1,-1)'$, $\mu_{4}=(3,-2)'$ for the ''well separated'' case and $\mu_{1} = (-1.5,0)'$, $\mu_{2}=(0.5,0.5)'$, $\mu_{3}=(1,-0.5)'$, $\mu_{4}=(3,-0.5)'$ for the ''overlapping'' case, respectively. The variance covariance matrices $\Sigma_k$ are chosen to be $0.25I$ for the separated case and $0.2I$ for the overlapping case for all $k=1,\ldots,4$. Figure \ref{fig:data_simulations} depicts two example datasets, representing the the two cases. In addition, we differentiate between a medium number of regressors (corresponding to the same true coefficient vectors as in section \ref{prior_study}) and a ''high sparsity'' case where 60 covariates that are not part of the data generating process are added to the covariate dataset, resulting in a total of 80 predictors in the model. Finally, we look at a case with a more complex sparsity structure as compared to section \ref{prior_study}. In this scenario, the first three predictors are only relevant to the first group, the second three predictors are only relevant to the second group and the third set of four predictors is only relevant to the third group. The last predictor is relevant for all groups. This setup requires the shrinkage priors to flexibly vary the amount of shrinkage by group. An overview of the four different simulation setups is given in Table \ref{tab:simoverview}.\\

\begin{table*}[!t]
\centering
\begin{threeparttable}
        \caption{Simulation Study II Overview}

\label{tab:simoverview}

 \footnotesize
        \setlength\tabcolsep{3pt}
        
        \renewcommand\arraystretch{1}

  \begin{tabular}{lccc}
    \toprule

          & \thead{Clusters} & \thead{No. of Regressors} & \thead{Sparsity Structure}\\

''Well Separated'' &    Separated & Medium & Simple \\ 
''Overlapping'' &    Overlapping & Medium & Simple \\ 
''High Sparsity'' &    Separated & Large & Simple \\ 
''Complex Sparsity'' &    Separated & Medium & Complex \\ 

\bottomrule
\end{tabular}
\end{threeparttable}
\end{table*}

For every setup, various summary statistics are computed. As before, RMSEs with respect to zero and non-zero coefficients as well as overall RMSEs and RMSEs with respect to the predicted probabilities are reported for the models with $K=4$. In addition, the misclassification rate of each model is computed. To assess the ability of the models to recover the true number of clusters, we run each simulation study for $K=(2,\ldots,6)$ and report plots of the average log Bayes factors\footnote{Where all models are assigned equal probabilities a priori.} with respect to the model with the true number of clusters $K=4$. Gibbs samplers using 2000 draws after a burn-in period of 5000 draws are implemented. Again, simulations are repeated 25 times and the resulting means of computed statistics across simulations are reported.

\begin{table*}[!h]
\centering
\begin{threeparttable}
        \caption{Simulation Study II Results}

\label{tab:simmain_c}

 \footnotesize
        \setlength\tabcolsep{3pt}
        
        \renewcommand\arraystretch{0.5}

  \begin{tabular}{ld{3.3}d{3.3}d{3.3}d{3.3}d{3.3}}
    \toprule

          & \thead{RMSE} & \thead{RMSE} & \thead{RMSE} & \thead{RMSE} & \thead{Miscl.} \\
        & \thead{(Zeros)} & \thead{(Non Zeros)} & \thead{(Overall)} & \thead{(P.P.)} & \thead{Rate} \\
\textbf{Well Separated}       &               &                   &                &             &      \\
Standard Prior &        2.09 & 1.41 & 1.68 & 1.17 & 0.02 \\ 
SSVS           &        1.09 & 1.03 & 1.05 & 1.06 & 0.01 \\ 
NG             &        1.00 & 1.00 & 1.00 & 1.00 & 0.01 \\ 
\midrule
\textbf{Overlapping}        &               &                   &                &             &      \\
Standard Prior &       3.05 & 2.22 & 2.52 & 1.37 & 0.07 \\  
SSVS           &       1.25 & 1.09 & 1.14 & 1.15 & 0.03 \\  
NG             &       1.00 & 1.00 & 1.00 & 1.00 & 0.03 \\  
\midrule
\textbf{High Sparsity}        &               &                   &                &             &      \\
Standard Prior &       9.10 & 5.94 & 7.68 & 1.27 & 0.06 \\  
SSVS           &       1.41 & 1.11 & 1.26 & 1.03 & 0.01 \\ 
NG             &       1.00 & 1.00 & 1.00 & 1.00 & 0.01 \\ 
\midrule
\textbf{Complex Sparsity}        &               &                   &                &             &      \\
Standard Prior &       1.92 & 1.86 & 1.89 & 1.17 & 0.02 \\ 
SSVS           &       1.02 & 1.17 & 1.08 & 1.04 & 0.01 \\ 
NG             &       1.00 & 1.00 & 1.00 & 1.00 & 0.01 \\  
\bottomrule
\end{tabular}
\begin{tablenotes}
\item\textit{Note:} The estimates correspond to the average value across 25 runs. RMSEs are reported relative to NG RMSEs.
\end{tablenotes}
\end{threeparttable}
\end{table*}

Table \ref{tab:simmain_c} reports the main simulation study results. On average, we find that the performance of the SSVS and NG prior is very similar in all simulation exercises. However, a slightly better performance of the NG prior can be found in some cases, especially in the high sparsity setting. Nevertheless, as mentioned before, RMSEs are small for SSVS and NG in absolute terms, suggesting both priors are in principal useful when selecting covariates in a mixture-of-experts framework. We do not report the estimation results for $\mu_k$ and $\Sigma_k$ as all three models perform extremely well in this regard, showing very similar results. The only issue that stands out is that model with the standard prior has a tendency to overestimate the variances. As pointed out by one of the reviewers, it might be illuminating to look at RMSEs that are separately computed by groups. This does not lead to any significant qualitative or quantitative variation in the results. Thus, RMSEs by group are not reported for brevity reasons. More detailed results are available from the author upon request.\\

\renewcommand\baselinestretch{1}
\begin{figure}[!t]
\centering
\subfloat[Standard Prior]{\includegraphics[width = 0.2\textwidth]{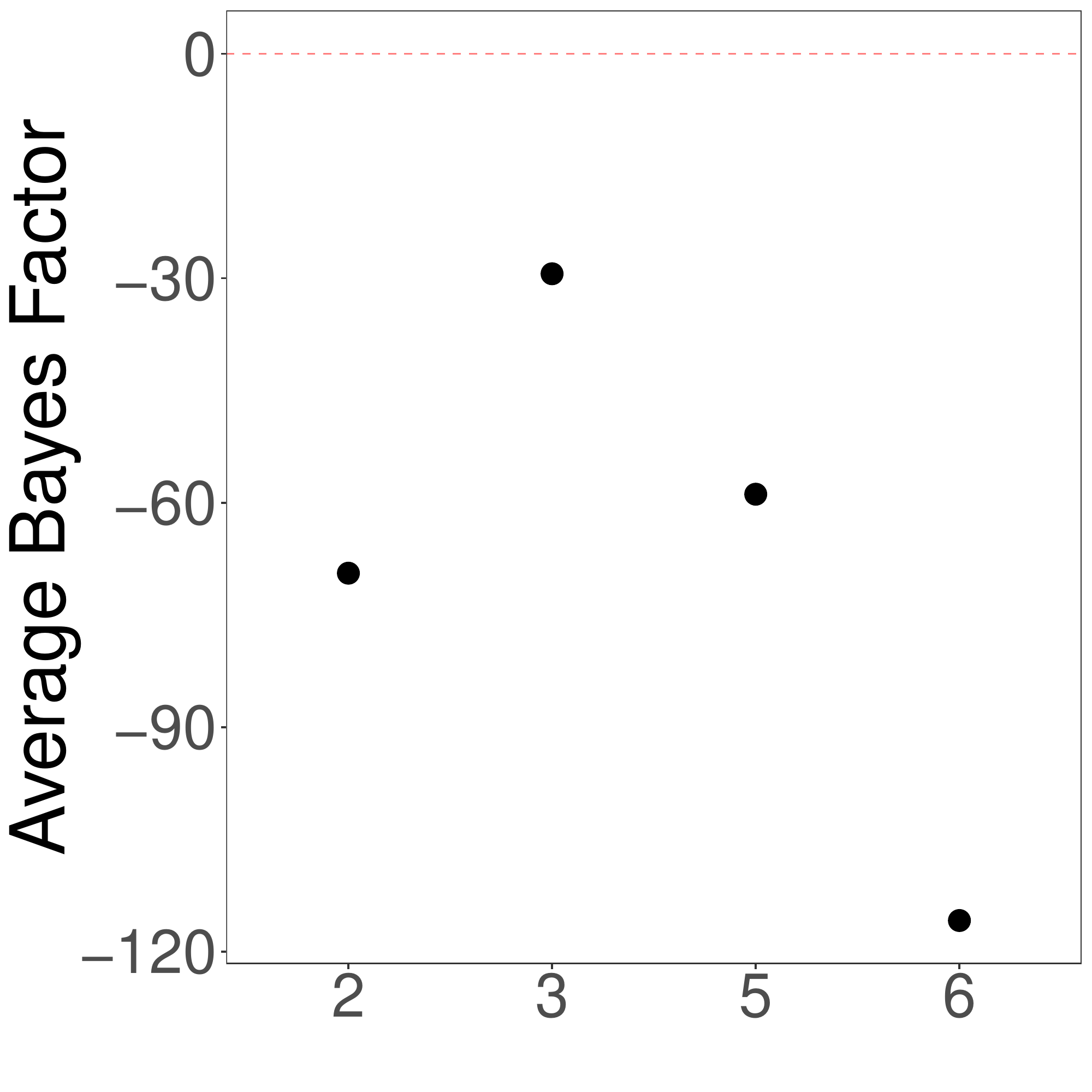}}
\subfloat[SSVS Prior]{\includegraphics[width = 0.2\textwidth]{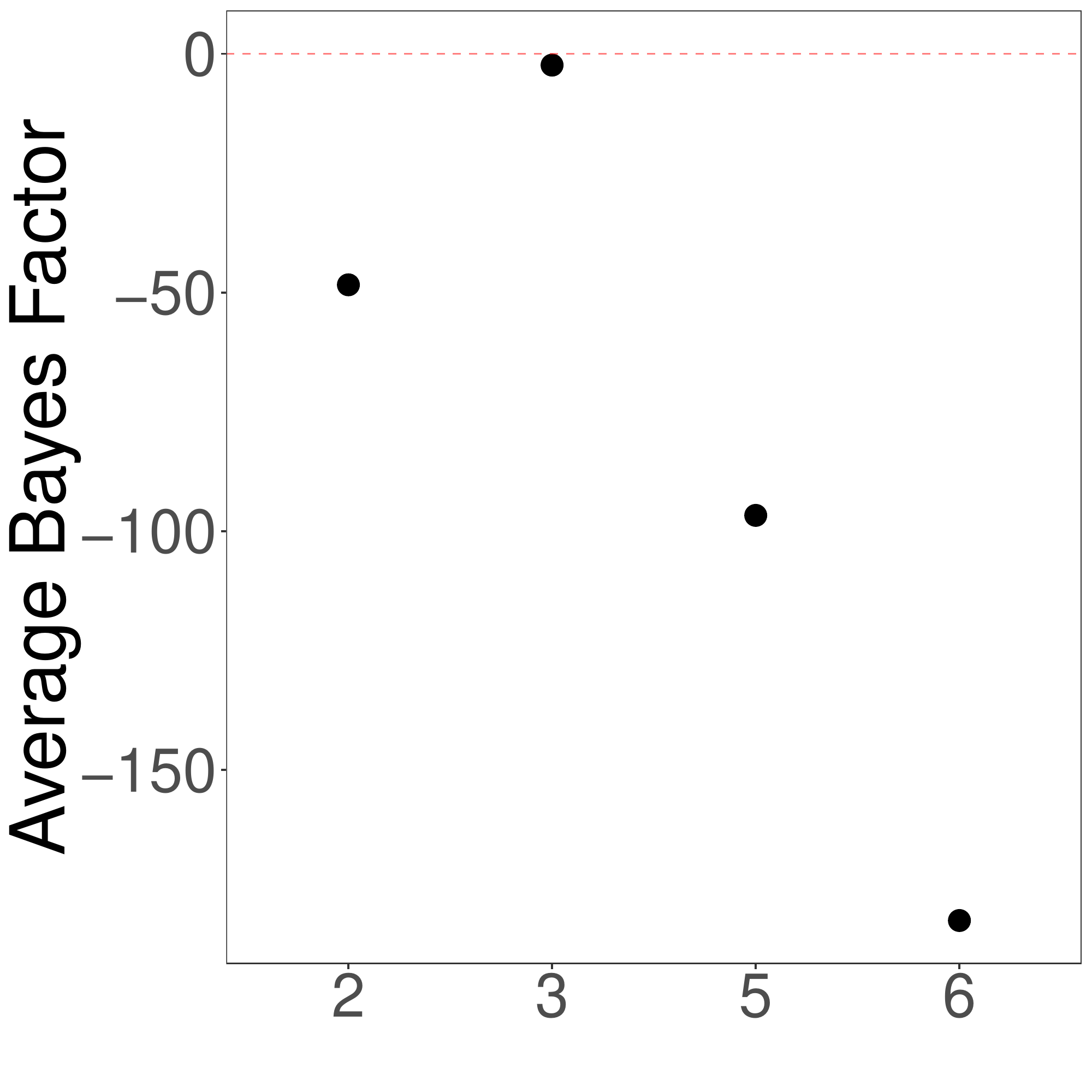}}
\subfloat[NG Prior]{\includegraphics[width = 0.2\textwidth]{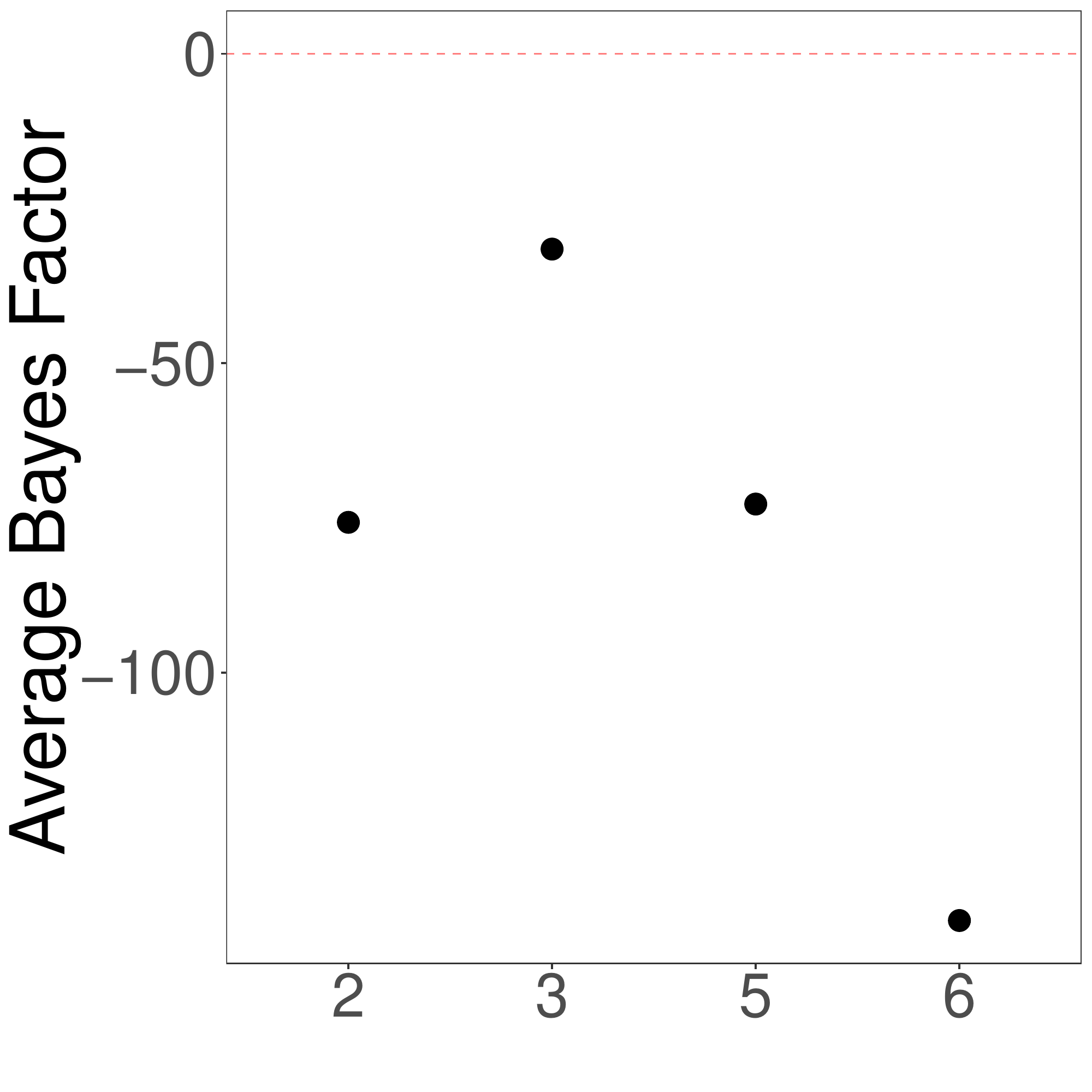}} 
\caption{Average Log Bayes Factors for different model setups (with dashed line at $0$) for well separated case. Reference Model is $K=4$.}
\label{fig:ml_sep}
\end{figure}
\begin{figure}[!t]
\centering
\subfloat[Standard Prior]{\includegraphics[width = 0.2\textwidth]{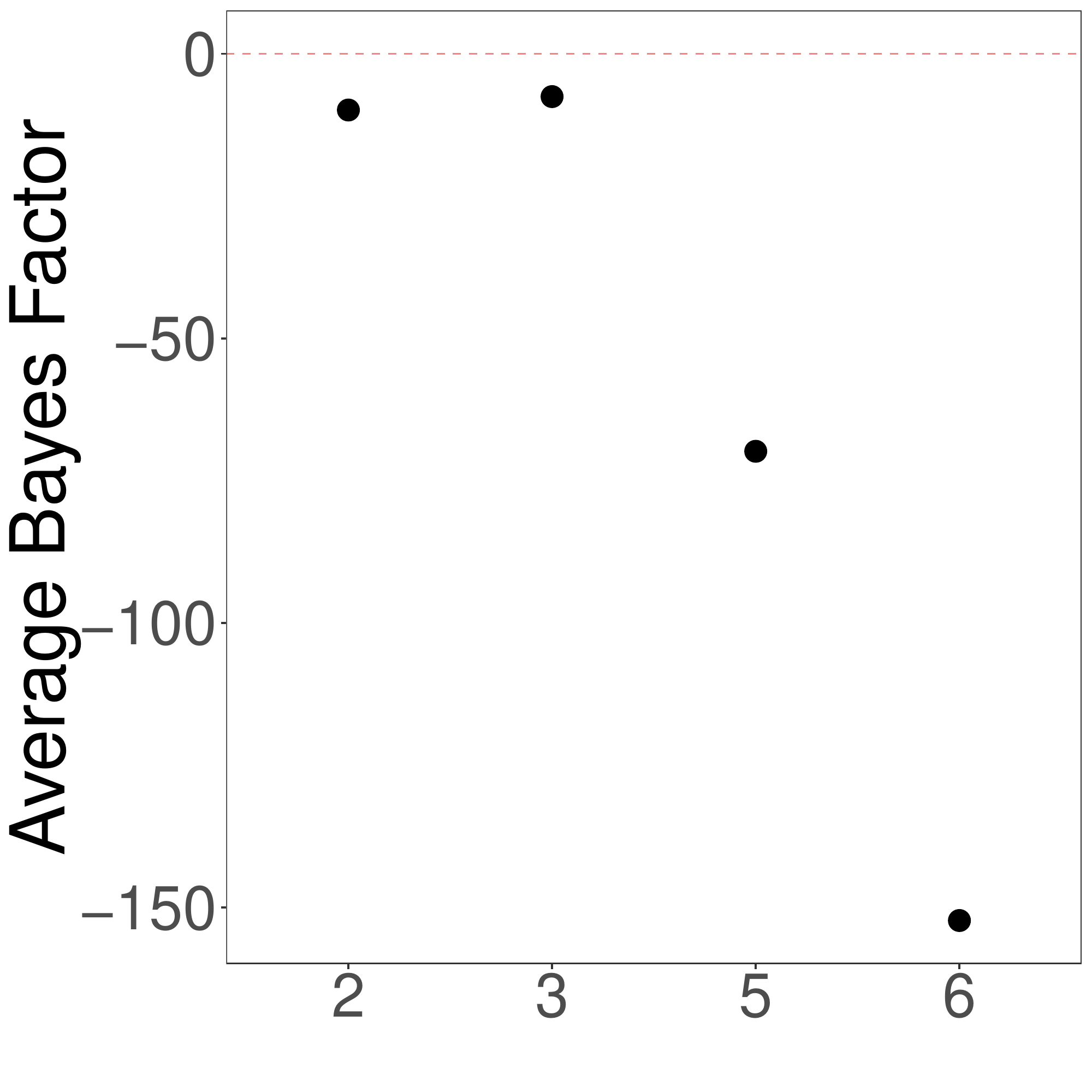}}
\subfloat[SSVS Prior]{\includegraphics[width = 0.2\textwidth]{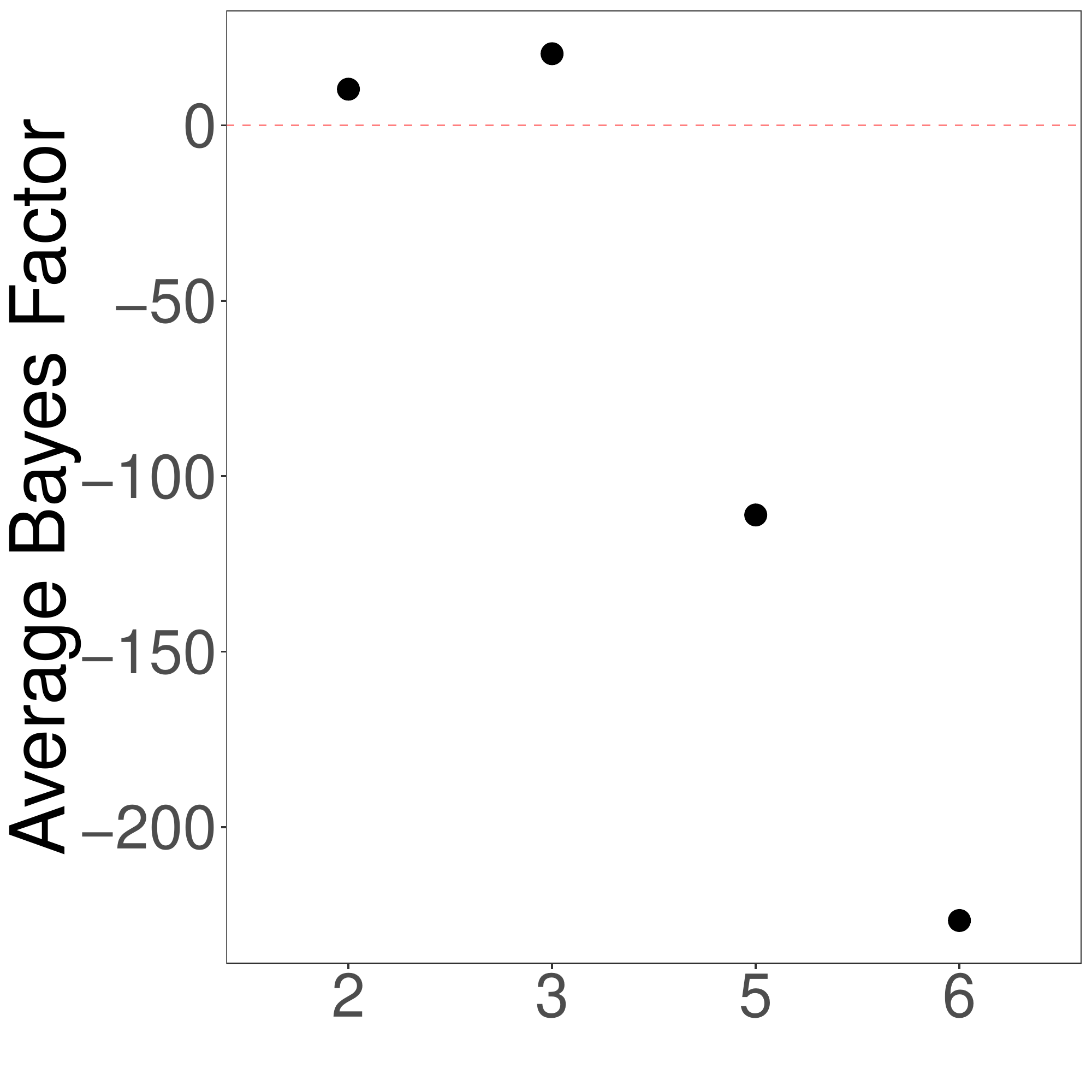}}
\subfloat[NG Prior]{\includegraphics[width = 0.2\textwidth]{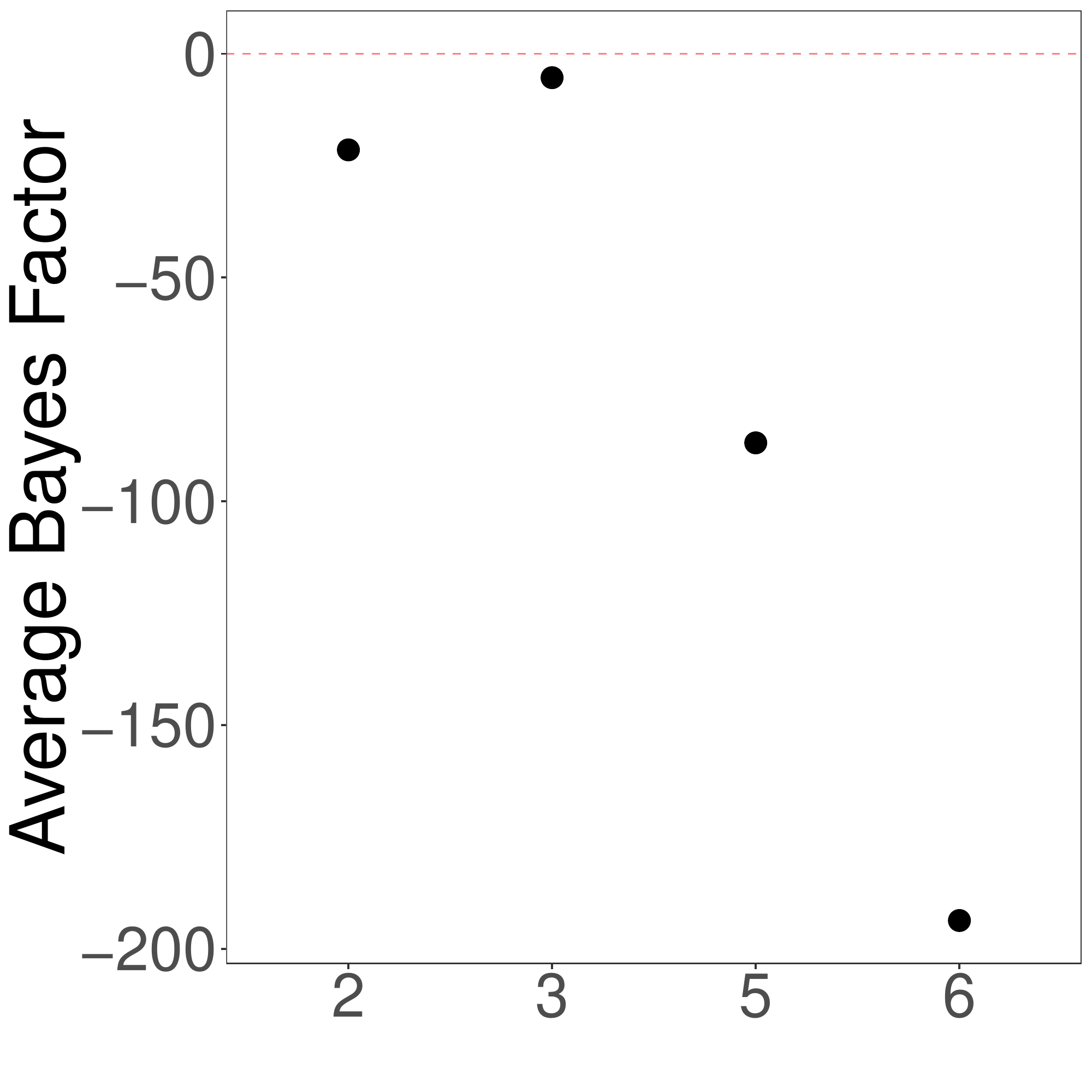}} 
\caption{Average Log Bayes Factors for different model setups (with dashed line at $0$) for overlapping case. Reference Model is $K=4$.}
\label{fig:ml_lap}
\end{figure}
\begin{figure}[!t]
\centering
\subfloat[Standard Prior]{\includegraphics[width = 0.2\textwidth]{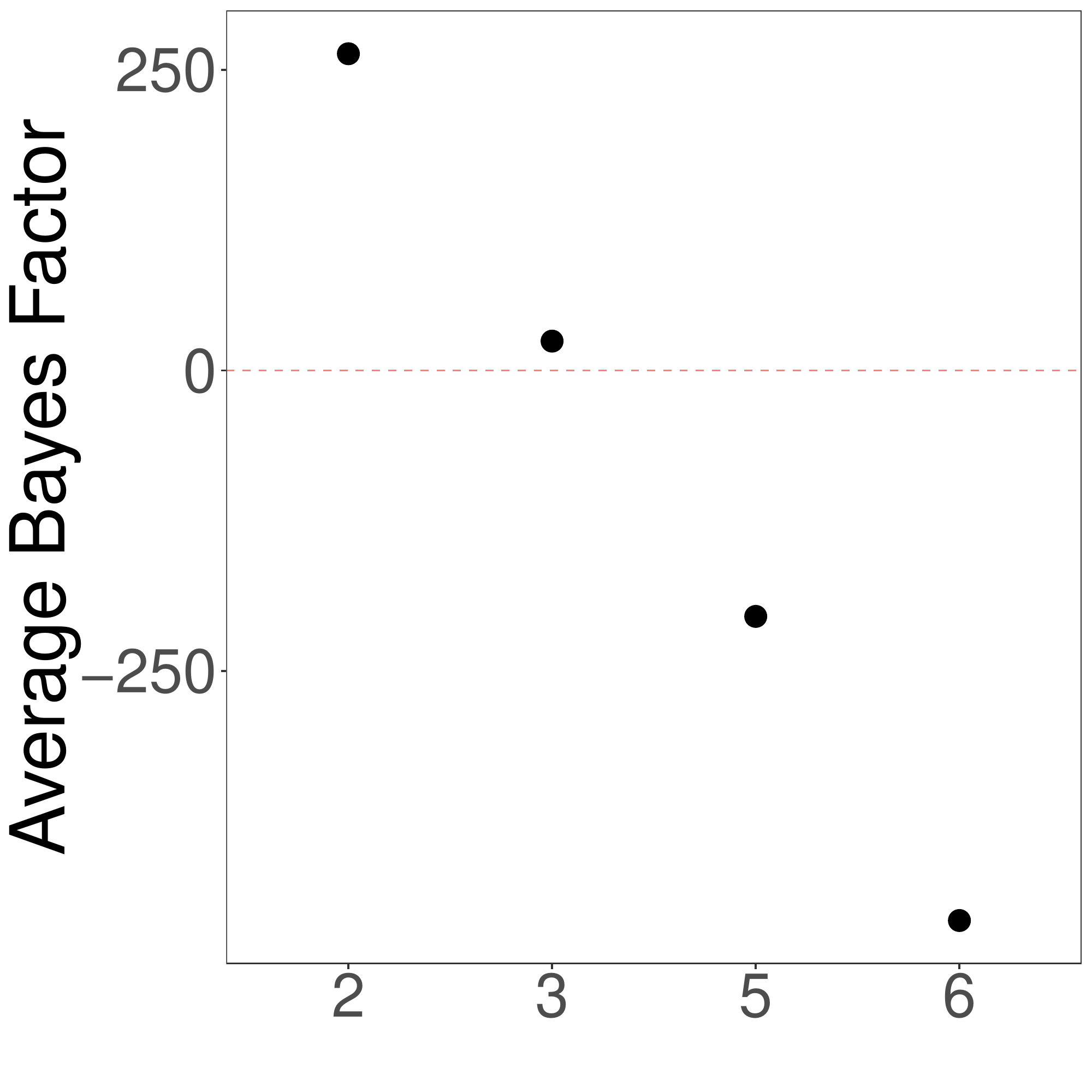}}
\subfloat[SSVS Prior]{\includegraphics[width = 0.2\textwidth]{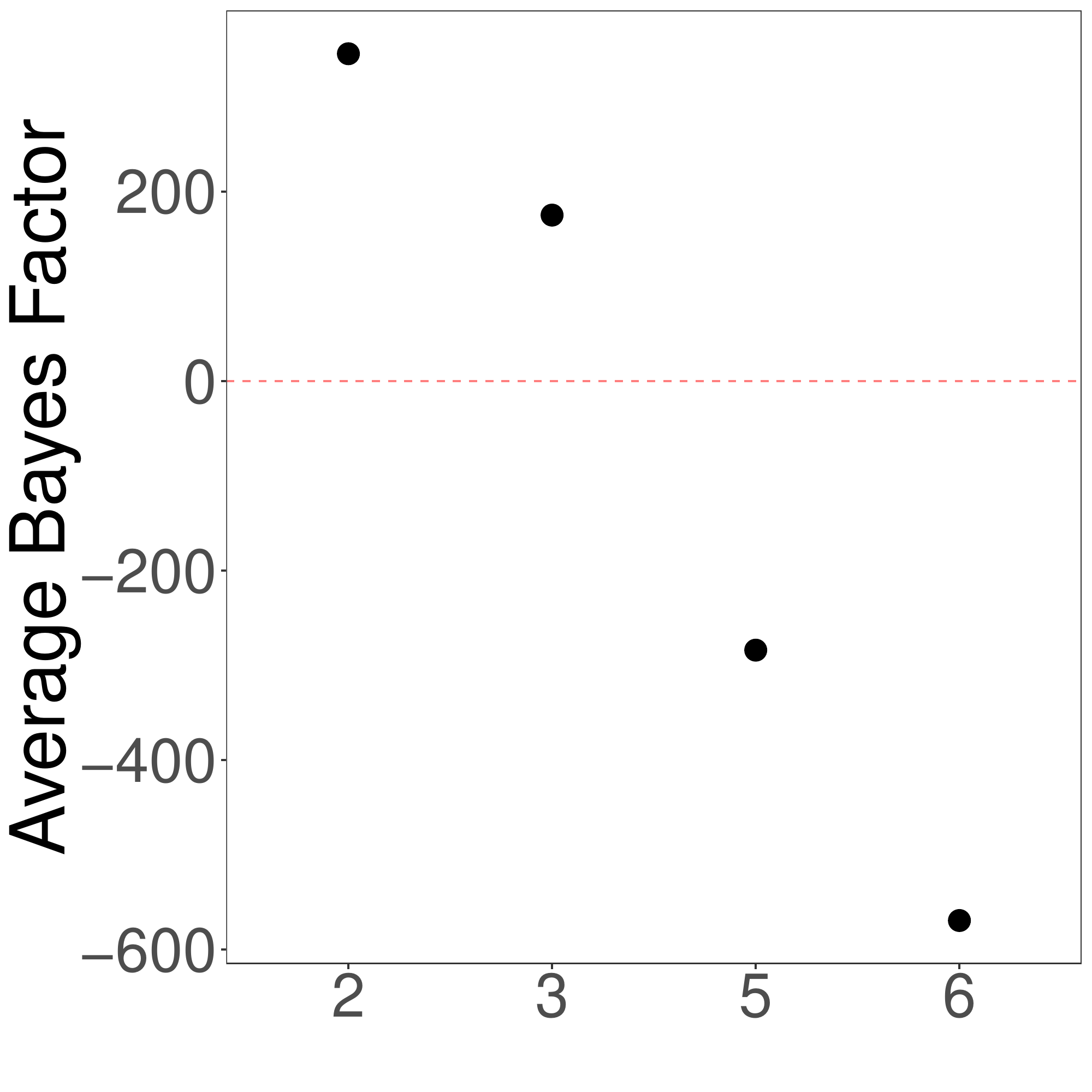}}
\subfloat[NG Prior]{\includegraphics[width = 0.2\textwidth]{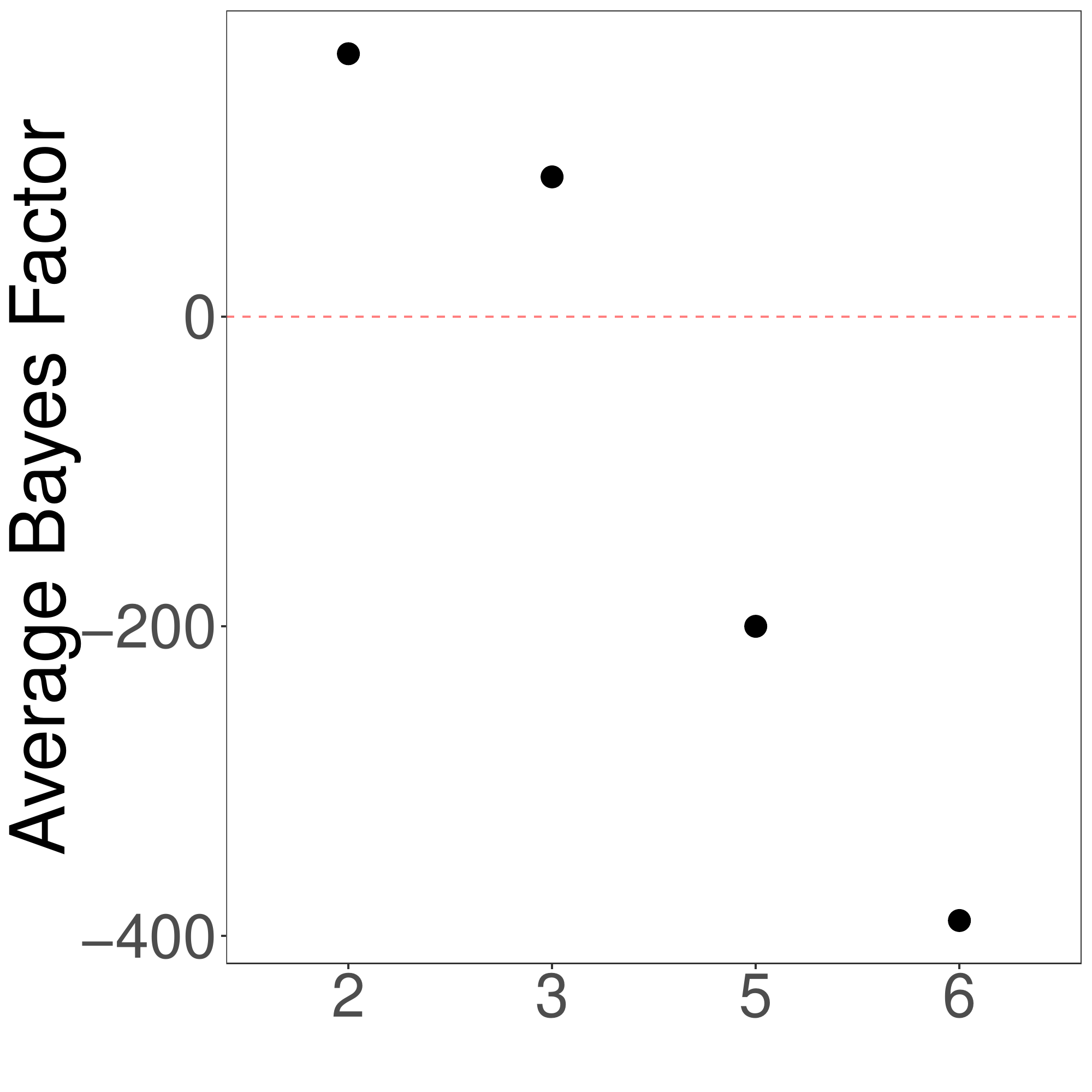}} 
\caption{Average Log Bayes Factors for different model setups (with dashed line at $0$) for high sparsity case. Reference Model is $K=4$.}
\label{fig:ml_spa}
\end{figure}
\begin{figure}[!t]
\centering
\subfloat[Standard Prior]{\includegraphics[width = 0.2\textwidth]{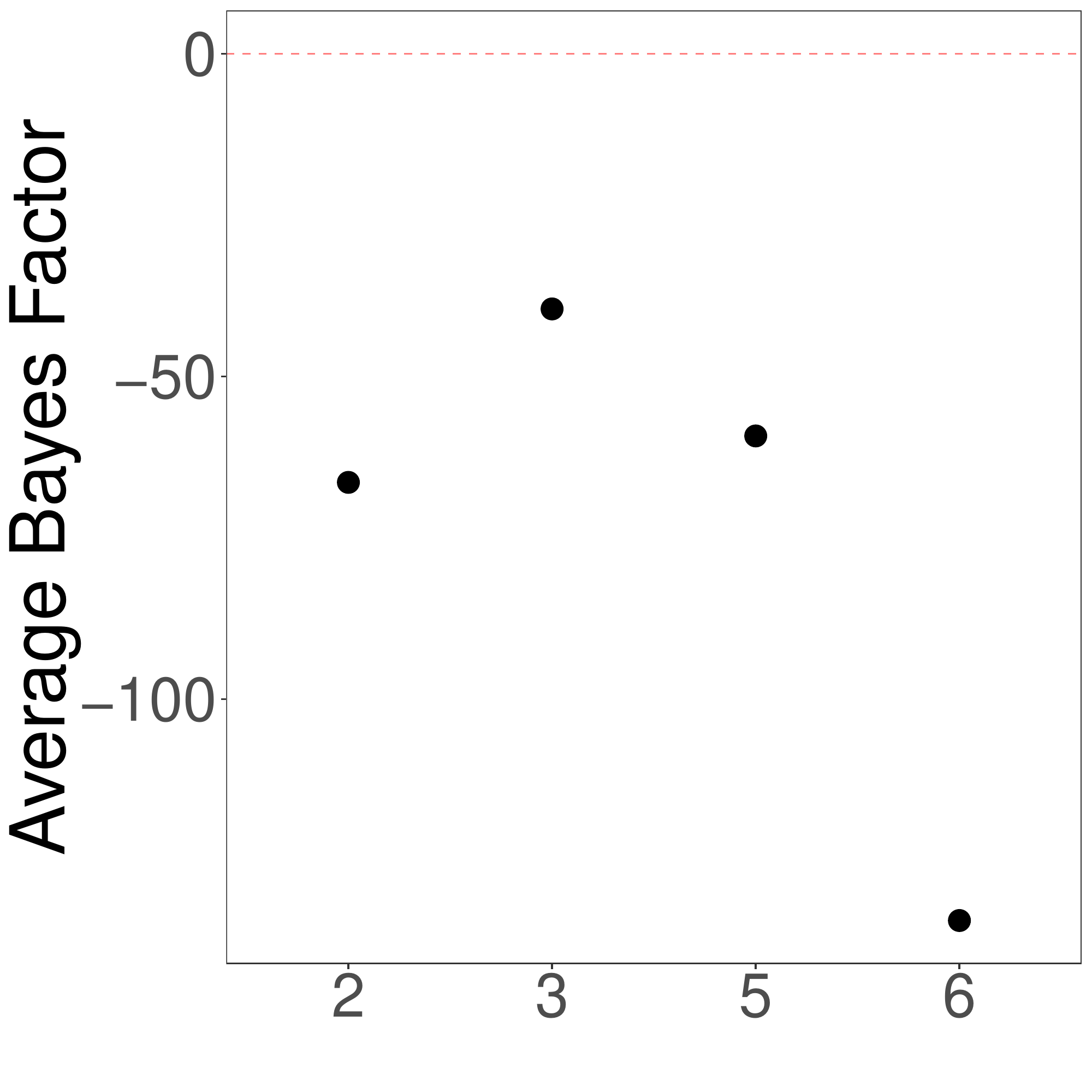}}
\subfloat[SSVS Prior]{\includegraphics[width = 0.2\textwidth]{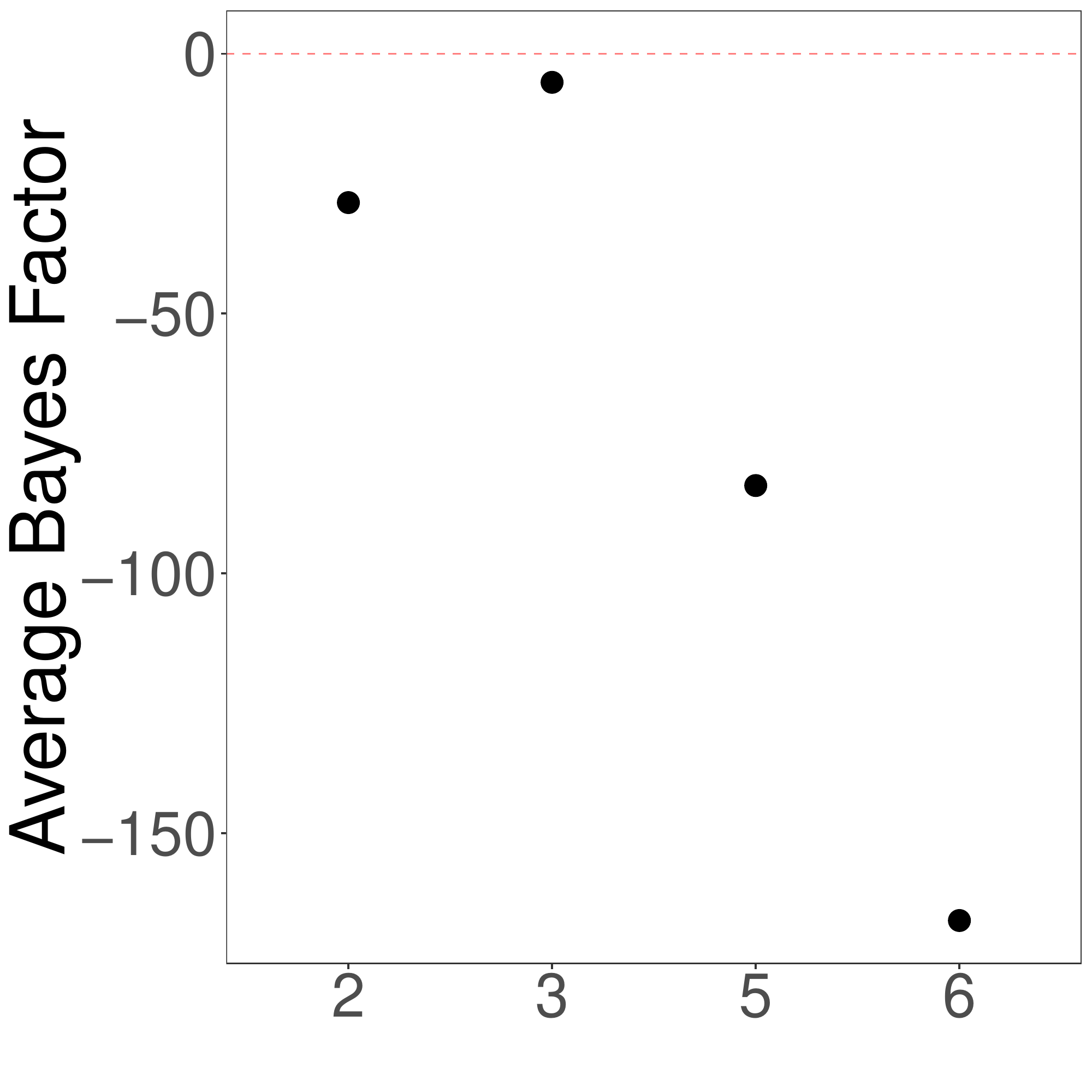}}
\subfloat[NG Prior]{\includegraphics[width = 0.2\textwidth]{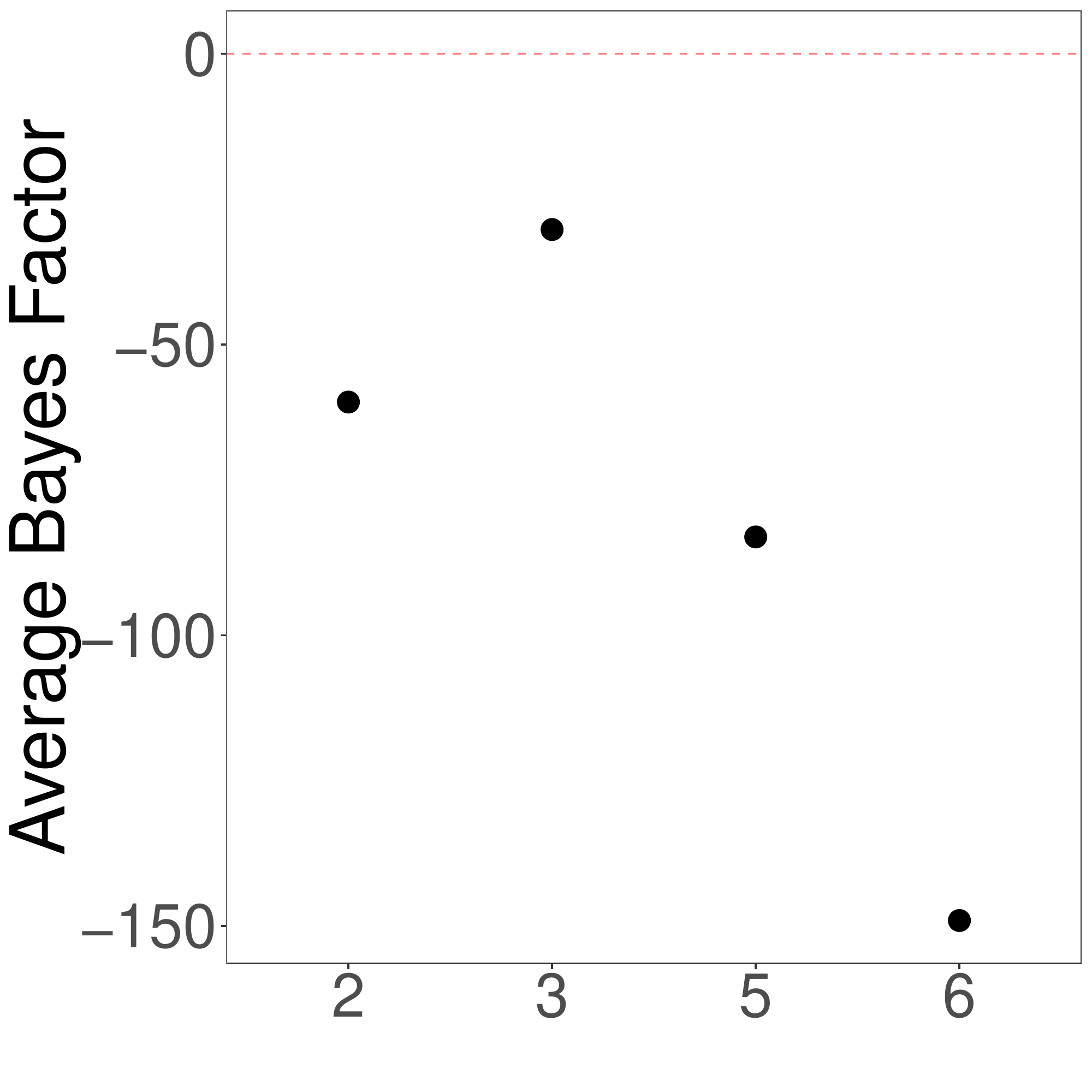}} 
\caption{Average Log Bayes Factors for different model setups (with dashed line at $0$) for complex sparsity case. Reference Model is $K=4$.}
\label{fig:ml_com}
\end{figure}
\renewcommand\baselinestretch{1.5}

Figures \ref{fig:ml_sep} - \ref{fig:ml_com} plot the average log Bayes factors relative to the true model with $K=4$ for the three priors. Positive values suggest that the respective model scored a higher marginal likelihood than the true model with $K=4$ and vice versa. In most cases, this model selection criterion suggests to choose $K=4$, with two exceptions: First, SSVS seems to have a slight tendency to favor models with a smaller number of clusters as compared to the other models. This leads to positive log Bayes factors for the models with $K=2$ and $K=3$ in the ''Overlapping'' simulation setup. Second, in the ''High Sparsity'' scenario, all log Bayes factors lean towards models with $K=2$ and $K=3$, suggesting influence of the number of predictors on the bridge sampling estimates. However, an in-depth examination of this issue is out of scope of this article and thus left for future research.\\
All in all, we conclude that the NG prior is a very useful alternative to SSVS in mixture-of-experts frameworks. Generally speaking, the results of the SSVS prior and the NG prior will be very similar, although there are some performance gains of the NG prior visible in terms of shrinkage as well as with respect to model selection issues.

\section{HIV information sources in Mozambique}

Mozambique is a country in Southeastern Africa that is considered one of the poorest and most underdeveloped countries in the world, scoring low in both economic and human development rankings. In the year 2008, Mozambique had the 8th highest HIV prevalence in the world with 1,600,000 people infected (11.6\% of the population) of whom around 990,000 were women and children. According to the Joint United Nations Programme on HIV/AIDS, there are around 590,000 HIV orphans living in Mozambique, 180,000 of whom are infected with the virus themselves, a large part due to mother-child transmission. 75\% of the infected population between the age of 15 and 19 is female. Moreover, a large gender disparity regarding the level of information on the disease can be observed. While around half of the male adolescent population has comprehensive knowledge on HIV, only 27.4\% of adolescent women have enough information to adjust their behaviour to protect themselves and their children according to the United Nations Children’s Fund. This disparity is suspected to be largely due to socioeconomic and sociocultural reasons, with the main drivers being traditional gender roles and religious involvement (\citealp{AGADJANIAN20051529}).\\
Consequently, it is crucial to isolate channels that can be used by the government and non-governmental organizations to disseminate vital information on HIV, especially to the female population. Informing females about HIV has proven not only to decrease the infection rate but also increase the economic and social independence of women (\citealp{audet2010sociocultural}). Our empirical example contributes to this relevant and important issue by clustering women in Mozambique into groups that are relatively homogenous with respect to their information sources on HIV, similar to \citet{dias2010modeling}. In addition, we use a large dataset of potential geographic and socioeconomic explanatory variables and isolate the most important factors that determine membership in those information clusters. The results may be used to derive for instance information campaign strategies for respective subgroups.

\subsection{Bayesian inference for mixtures of Bernoulli distributions}

We use a set of binary variables that indicates whether a particular woman uses a specific source to gather information on HIV or not. A convenient choice of mixture distribution is the Bernoulli distribution, which proves useful when clustering binary vectors (see for example the vast literature on market segmentation; \citealp{wedel2012market}).\\
Let $y_i = (y_{i,1},\ldots,y_{i,J})$ be a $J$-dimensional vector of 0s and 1s that describe the HIV information sources used by woman $i = 1,\ldots,N$. Assume that this vector is the realization of a binary multivariate random variable $Y = (Y_1,\ldots,Y_J)$. Now suppose there exist $K$ groups in the population that cause differences in occurence probabilities $\gamma_{k,j} = \text{Pr}(Y_j = 1 |S_i = k)$ in $K$ different classes for $J$ different binary variables. $S_i$ is the latent class indicator of woman $i$.  We can rewrite Eq. \ref{MOE} where $y_i$ follows the mixture distribution

\begin{equation}
f(y_i ~|~ x_i) = \sum_{k=1}^K \ \eta_k(x_i) ~ \prod_{j=1}^J \gamma_{k,j}^{y_{i,j}}~(1-\gamma_{k,j})^{1-y_{i,j}}.
\end{equation}

The $K$ components correspond to the latent classes in the population. This model is widely used in various research fields, starting as early as \citet{lazarsfeld1959latent}. For details, see \citet[Ch. 9.5]{fruhwirth2006}. We assume that all probabilities $\gamma_{k,j}$ are a priori independent and specificy a beta prior of the form

\begin{equation}
    \gamma_{k,j} \sim B(a_{0,j},b_{0,j})
\end{equation}

and derive the posterior distribution conditional on the latent class indicators $S_i$, given by
\vspace{-0.1cm}
\begin{equation}
    \gamma_{k,j}|S,y \sim B(a_{0,j} + N_{k,j}, b_{0,j} + N_k - N_{k,j})
\end{equation}

where 

\begin{equation}
    \begin{split}
        N_k &= \sum_{i=1}^N \mathds{1}(S_i = k)\\
        N_{k,j} &= \sum_{i=1}^N y_{i,j} \mathds{1}(S_i = k).
    \end{split}
\end{equation}

\subsection{Data Description}

We apply the proposed model to data compiled from the Demographics and Health survey (DHS) for Mozambique from 2003. The DHS is a nationally representative household survey on a wide range of topics, including HIV information sources and various socioeconomic, geographic and health related variables.\\
The dataset includes information on 11,922 women. Ten different information sources are used to cluster these women into groups and a set of around 40 external covariates enters the model to explain class membership. These variables cover socioeconomic characteristics like age and education, region of residence, relationship status and sexual behavior as well as poverty related measures and dwelling characteristics. Table \ref{dataset} provides a detailed overview of the candidate explanatory variables.
\renewcommand\baselinestretch{1}
\begin{table*}[t]
\centering
\begin{threeparttable}
\caption{Dataset Overview.}
\footnotesize
        \renewcommand\arraystretch{0.6}
\label{dataset}
\begin{tabular}{p{4.3cm}p{5.2cm}p{5cm}cr}
\toprule
\textbf{Household Level Variables}           & Description                                                                             & Comments                                                                  \\
\midrule
\textbf{Dwelling Characteristics}      &                                                                                         &                                                                           \\
Household Size                         & Number of persons living in household.                                                    &                                                                           \\
Electricity                     & Household has access to electricity.                                              &                                                                           \\
Toilet               &            Household has access to a flush toilet.                                    &                                                                           \\
Phone                  & Household has a telephone.                     &                                                                           \\
Bicycle                          & Household has access to a bicycle.                      &                                                                           \\
Motorbike                          & Household has access to a motorbike.                      &                                                                           \\
Car                          & Household has access to a car.                      &                                                                           \\
Wealth                     & Position in the wealth distribution.                                         & Dummies for five wealth levels.               \\
                              &                                                                                         &                                                                           \\
\textbf{Geography}                     &                                                                                         &                                                                           \\
Province                          & The province the household is located in.                                                            & Dummies for the ten provinces of Mozambique.                        \\
Urbanisation                  & Degree of Urbanisation.                                                                 & Dummy for rural areas.
                              &                                                                                         &                                                                           \\
\midrule
\textbf{Personal Level Variables}      &                                                                                         &          
                                                      \\ \midrule
\textbf{Socioeconomic Characteristics} &                                                                                         &                                                                           \\
Relationship Status                       & Marital status of the women.                                                       & Dummies for married, single, partnership, widowed, living separated and divorced.                                             \\
Age                           & Age of the women.                                                                  &                                                                           \\
Sexual Activity                          & Number of sex partners in last 12 months.                                                                &                                                                        \\
Religion                   & Religion of the women.                                                     & Dummies for Catholic, Protestant, Muslim, African Zionist, Spiritualist, other and no religion.                       \\
Education                           & Years of schooling.                                                                &                                                                         \\
Literacy                           & Classification of the reading ability of the women.                                                                &  Dummies for literate, reading problems and illiterate.                                                                       \\
Employment Status                           & Employment status of the women.                                                                &  Dummy for unemployed.                                                                       \\

                              &                                                                                         &                                                                           \\

\textbf{HIV Information Sources} &                                                                                         &                                                                           \\
Information Source            &   Type of HIV information sources the women use.  &  Dummies for Radio, TV, Newspapers / Magazines, Posters, Clinic / Healthworker, Church, School, Community Meetings, Friends / Relatives and Working Place.\\
\bottomrule
\end{tabular}
\begin{tablenotes}
\item\footnotesize
\end{tablenotes}
\end{threeparttable}
\end{table*}
\renewcommand\baselinestretch{1.5}
\subsection{Results}

\begin{figure}[h!]
    \centering
    \includegraphics[width=0.2\textwidth]{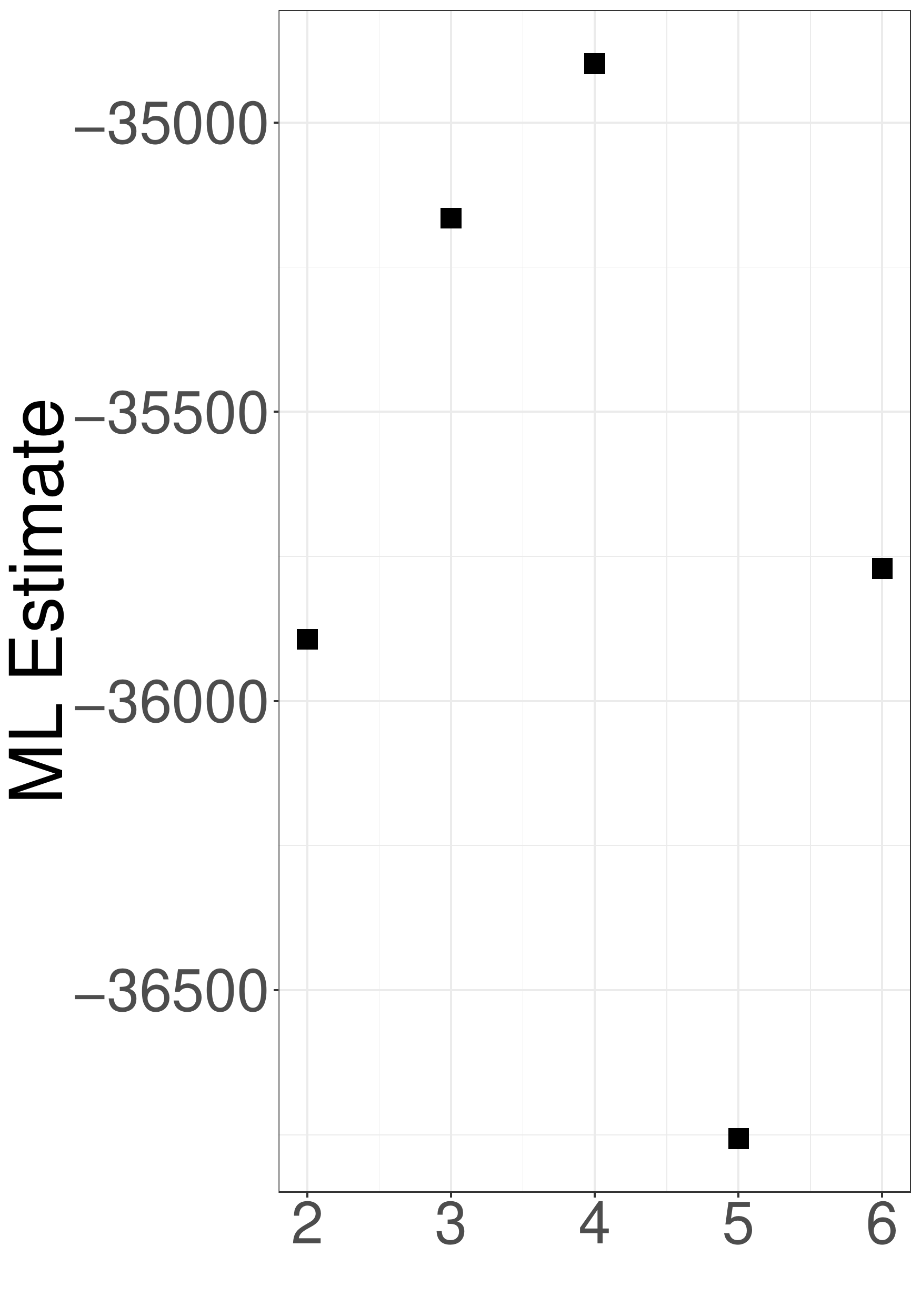}
    \caption{Estimates of log marginal likelihoods.}
    \label{fig:ml}
\end{figure}

\begin{figure}[h!]
    \centering
    \includegraphics[width=\textwidth]{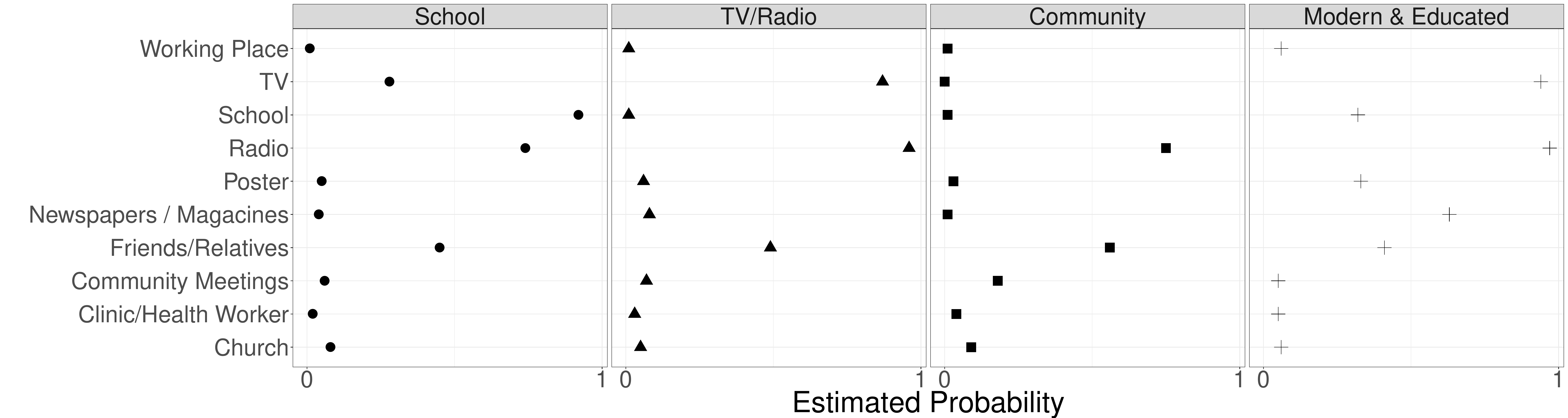}
    \caption{Information source estimates for each cluster.}
    \label{parameters}
\end{figure}

We estimate the model with a NG shrinkage prior for different values of $K$ and compare the resulting models using the marginal likelihood estimates obtained via bridge sampling.\footnote{The model has been implemented in R (\citealp{R}). Computational time for $K=4$ is around 50 minutes for 5000 draws after a burn in period of 1000 draws on an Intel i7 @ 2.4 GHZ.} We choose the model that maximizes the marginal likelihood. The bridge sampling estimates of the log marginal likelihood for $K=2,\ldots,6$ are provided in Figure \ref{fig:ml}. The model with $K=4$ scores highest and is therefore discussed below.\footnote{This is of course not the only way to proceed here. Especially in a development context, other, more informal model selection criteria that take into account long term campaigning strategies or financial constraints may be employed. For example, the groups "Modern \& Educated" and "TV/Radio" could be merged as they are both have a distinctive dependence on TV. However, in this paper the statistical possibilities of the proposed model are emphasized and hence we make use of the purely statistical approach.}
The estimates for $\gamma_{k,j}$ are presented in Figure \ref{parameters}. The uncertainty surrounding these estimates is usually extremely small. At first glance we find that the radio as well as friends and relatives seem to be important information sources for all groups. For the purpose of further interpretation of the model results, we name the groups with respect to their most distinctive HIV information source as described below.

%\begin{table*}[h]
%\centering
%\footnotesize
%\setlength{\tabcolsep}{8pt}
%        \renewcommand\arraystretch{0.5}
%\begin{threeparttable}
%\caption{Information source estimates for each cluster.}
%\label{parameters}
%\begin{tabular}{lccccccccc}
%\toprule
%Source & Modern \& Educated  & TV & School & Church/Community & Friends/Relatives   \\
%\midrule
%Radio & 0.97 & 0.96 & 0.74 & 0.64 & 0.81 \\ 
%TV & 0.91 & 0.97 & 0.30 & 0.01 & 0.00 \\ 
%Newspapers / Magazines & 0.60 & 0.01 & 0.04 & 0.01 & 0.01 \\ 
%Poster & 0.32 & 0.02 & 0.05 & 0.02 & 0.03 \\ 
%Clinic/Health Worker & 0.06 & 0.02 & 0.02 & 0.06 & 0.03 \\ 
%Church & 0.06 & 0.05 & 0.08 & 0.26 & 0.01 \\ 
%School & 0.25 & 0.00 & 0.95 & 0.02 & 0.00 \\ 
%Community Meetings & 0.05 & 0.07 & 0.06 & 0.35 & 0.11 \\ 
%Friends/Relatives & 0.41 & 0.49 & 0.46 & 0.46 & 0.60 \\ 
%Working Place & 0.05 & 0.01 & 0.01 & 0.02 & 0.00 \\ 
%\midrule
%Group size & 0.102 & 0.095 & 0.062 & 0.227 & 0.514 \\ 
%\bottomrule
%\end{tabular}
%\begin{tablenotes}
%\footnotesize
%\item\textit{Note:} Estimated values correspond to the posterior means of $\gamma_{k,j}$ and to the actual group sizes.
%\end{tablenotes}
%\end{threeparttable}
%\end{table*}

\begin{figure}[h!]
    \centering
    \includegraphics[width=0.95\textwidth]{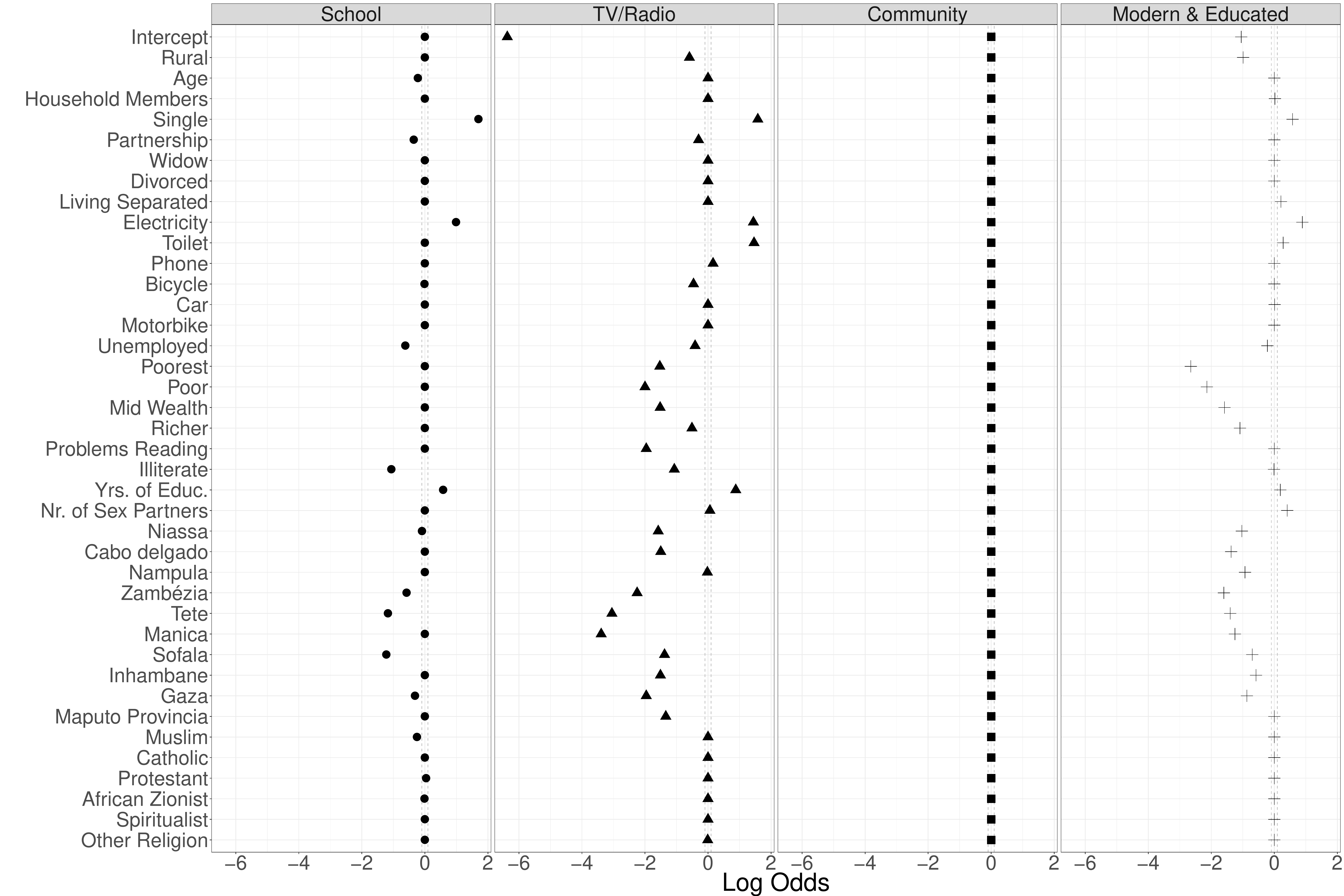}
    \caption{Posterior medians for multinomial logistic regression coefficients (Baseline: "Friends/Relatives").}
    \label{fig:my_label}
\end{figure}

Around 8\% of the female population use modern information sources such as television, newspapers and posters. In addition, this group obtains a relatively high amount of information from schools. Thus, we label this group as ''Modern \& Educated''. A somewhat larger group (around 13\% of the population) relies mostly on TV but is highly unlikely to inform themselves in schools. Hence, we name this group ''TV/Radio''.  The third group, ''School'', which comprises around 6\% of the female population of Mozambique, relies heavily on schools for obtaining information on HIV. The baseline group ''Community'' (73\%) has an above average dependence on friends and relatives, community meetings and local churches in terms of information on the disease.\\
Figure \ref{fig:my_label} provides a plot of the point estimates of the logit coefficients.

When interpreting the multinomial logit coefficients, one has to keep in mind that the effects are always interpreted with respect to a baseline group. For convenience in the estimation process, we choose the largest group as baseline group (''Community''). In terms of the other categorical variables, the ''baseline woman'' is residing in Maputo City, has no religion, is married and is a member of the richest wealth group.\\
Strong effects of the wealth distribution on the probability of being a member of the ''Modern \& Educated'' and ''TV/Radio'' group are observable. These groups also share an above average probability of having access to a flush toilet and electricity. In addition, it is relatively unlikely that a woman lives in the countryside and is a member of one of those groups. These findings are in line with what theory suggests in a poverty plagued country like Mozambique.\\
Unmarried women with above average education are also likely to rely on schools as HIV information sources. This seems puzzling, as female education is a primary development issue in various African countries. However, one should keep in mind that this group is extremely small and comprises just above 6\% of the female population. Interestingly, wealth variables seem to be not strongly correlated with the probability of being a member of the ''School'' group as compared to the other groups. However, we see that geographic variables determine prior class membership for this group, implying that we can find spatially clustered communities with above average female educational attainment throughout specific provinces.\\
These results are particularly relevant to policy makers. Important insights that can be derived are, for instance, that HIV information campaigns that are targeted on disseminating educational materials via schools are likely to be more effective in Maputo City as compared to Sofala. It also might be a good idea to target folders that are distributed in schools towards single women as opposed to married women. However, these are mere examples. A detailed discussion of the policy implications of the results is out of scope of this article.\\

\section{Concluding Remarks}

Finite mixture models are a commonly used tool for model based clustering and density estimation. They can be extended to mixture-of-experts models, allowing to use information from several covariates when clustering dependent variables of arbitrary form. We propose the usage of continous shrinkage priors to find robust predictors of class membership in this context. This enables us to simultaneously identify underyling groups in a population, cluster said observations into these groups and find the important predictors of being a group member. In particular, we suggest a combination of the normal gamma prior (\citealp{griffin2010inference}) and the P\'{o}lya-Gamma sampler (\citealp{polson2013bayesian}) for implicit variable selection in a multinomial logistic regression that is used to model prior class membership.\\
This setup solves the issue of model uncertainty that arises in this context and reduces the sensitivity of the model with respect to included variables. The proposed framework slightly outperforms related approaches and makes more precise clustering  in setups with a large number of predictor variables possible.\\
We illustrate the model in a real data application where we apply the model with a mixture of Bernoulli distributions to HIV information sources of women in Mozambique. Model selection is based on the bridge sampling estimate of the marginal likelihood. We find four clusters of women who are relatively homogenous with respect to their HIV information sources. Somewhat unsurprising, we find that wealth plays an important role in the access to information on HIV. Moreover, geographical patterns of information seeking behavior seem to be prevalent.\\
Further research may be pointed into the direction of comparing the performance of different shrinkage priors in this context in a more detailed way as seen in \citet{sfswagner}. One promising candidate is for example the Dirichlet-Laplace prior from \citet{bhattacharya2015dirichlet}. It might also be possible to extend various other Bayesian variable selection methods to mixture-of-experts frameworks, for example Bayesian compression (\citealp{guhaniyogi2015bayesian}) or its extension using targeted random projections (\citealp{mukhopadhyay2017targeted}). Another interesting problem is how to apply the idea of shrinkage introduced through the prior class membership weights (e.g. \citealp{malsiner2016model}) for model selection purposes into a mixture-of-experts framework. Also, the evaluation of the forecasting performance of the model was not attempted in this article and is left for further research.

%\begin{acknowledgements}
%\end{acknowledgements}

\bibliographystyle{spbasic}      % basic style, author-year citations
\bibliography{lit}   % name your BibTeX data base

\clearpage

\appendix
\normalsize
\section{Choice of Hyperparameters \& Priors}
\label{hyperparams}

To enable estimation, it is necessary to choose values for various (hyper-)parameters appearing in the model setup. We set $c0 = c1 = 0.01$ to obtain a rather uninformative prior distribution as for instance seen in \cite{huber2017adaptive}. However, the only choice of parameter that influences inference is in fact the choice of $\theta$. Using values close to zero induces rather heavy shrinkage whereas higher values correspond to significantly less shrinkage. As the motivation of the empirical example is to isolate robust determinants of class membership and not to find precise point estimates, we set $\theta$ to the comparably small value of $0.05$ and take the risk of overshrinking some parameters. We set $\theta$ to $0.1$ in the simulation study. A thorough discussion of the choice and influence of $\theta$ can be found in \cite{bitto2016achieving}.

\section{Bridge sampling in mixture-of-experts Models}
\label{bridge}
A first step in computing the bridge sampling estimate for the proposed model is to construct an importance density that approximates the modes of the posterior density. As the posterior density of a mixture model will have multiple modes, this problem turns out to be challenging. As one of the proposed model's benefits is that all posterior distributions are available in closed form, we can make use of the unsupervised importance density construction that has been suggested by \cite{fruhwirth1995bayesian} and extended by \cite{fruhwirth2004estimating}. The idea is to choose a random subsample of $S$ posterior densities from the $M$ available permutated MCMC draws and use them to automatically construct the importance density. As we use random permutation sampling, this importance density will be multimodal as well.\\ 
In practical terms, it is necessary to save the posterior distribution parameters of $S$ randomly selected MCMC draws during the sampling process. Note that this implies that the $S$ saved parameters of the posterior distributions are not part of the ex post identification procedure. If one has chosen a suitable number of importance densities $S$ and number of draws from the importance density $L$, we can proceed and draw from the importance density. The idea is to draw from a uniform mixture of $S$ posterior densities. We implement this step as follows. For $l = 1,\ldots,L$:

   \begin{enumerate}
       \item Choose a random index out of the $1,\ldots,S$ saved posterior density parameters.
       \item For $k = 1,\ldots,K$, generate one draw from the posterior densities with the parameters that have been randomly chosen in the previous step. Iterate this procedure.
   \end{enumerate}

The obtained $M$ MCMC draws and $L$ importance density draws can be used in the recursive iteration scheme that has been described in Section \ref{modelselection}.\\
To run the iterative process, several likelihoods have to be evaluated:

\begin{enumerate}
    \item Evaluate the importance density draws in the prior densities.
    \item Evaluate the importance density draws in the importance density.
    \item Evaluate the mixture likelihood using the importance density draws.
    \item Evaluate the MCMC draws in the prior densities.
    \item Evaluate the MCMC draws in the importance density.
    \item Evaluate the mixture likelihood using the MCMC draws.
\end{enumerate}

For a detailed and more formal description for this procedure, see \cite{fruhwirth2004estimating} and \cite{modelselection}.

\section{Numerical Stability of Bridge Sampling Estimate}
\label{numeric}
\renewcommand\baselinestretch{1}

Depending on the sample size $N$, the number of MCMC draws $M$ and both the number of densities chosen to construct the importance density $S$ and the number of importance density draws $L$, the vectors and matrices that result from evaluating the likelihoods will be large. Hence, the evaluated log-likelihoods may be small in absolute values (e.g. $-0.1$), but summing over a large number of log likelihoods and exponentiating this sum is prone to numerical underflow. Therefore, we suggest a specific evaluation scheme that has proved numerically stable in our computations. It is based on the idea that we can rewrite the log of the bridge sampling estimate of the marginal likelihood as a double log sum of exponentials. Then we can exploit the following identity:

\begin{equation*}
     \text{log}(\sum_ie^{x_i}) = \text{max}(x_i) + \text{log} (\sum_i e^{x_i - max(x_i)}).
\end{equation*}

This LogSumExp function can be used to generate an exact and numerically stable estimate of the logarithm of the sum of exponential terms. To employ this function in the bridge sampling procedure, we rewrite the equation of the bridge sampling estimate as follows.
\begin{landscape}

\begin{equation*}
  \begin{aligned}
  \text{log}\Big(\hat{p}_{BS,t+1}\Big) &= \text{log}\Big(\frac{L^{-1}\sum_{l=1}^L \frac{p^{\star}_l}{Lq_l + Mp^{\star}_l / \hat{p}_{BS,t}}
 }{M^{-1}\sum_{m=1}^M  \frac{q_m}{Lq_m + Mp^{\star}_m / \hat{p}_{BS,t}}}\Big)\\[8pt]
  &= -\text{log}(L) + \text{log}\Big(\sum_{l=1}^L \text{exp}\big(\text{log}(\frac{p^{\star}_l}{Lq_l + Mp^{\star}_l / \hat{p}_{BS,t}})\big)\Big) - \Big[-\text{log}(M) + \text{log}\Big(\sum_{m=1}^M  \text{exp}\big(\text{log}(\frac{q_m}{Lq_m + Mp^{\star}_m / \hat{p}_{BS,t}})\big)\Big)\Big]\\[8pt]
 &= -\text{log}(L) + \text{log}\Big(\sum_{l=1}^L \text{exp}\big(\text{log}({p^{\star}_l})-\text{log}({Lq_l + Mp^{\star}_l / \hat{p}_{BS,t}})\big)\Big) - \Big[-\text{log}(M) + \text{log}\Big(\sum_{m=1}^M  \text{exp}\big(\text{log}({q_m})-\text{log}({Lq_m + Mp^{\star}_m / \hat{p}_{BS,t}})\big)\Big)\Big]\\[8pt]
  &= -\text{log}(L) + \text{log}\Big(\sum_{l=1}^L \text{exp}\big(\text{log}({p^{\star}_l})-\text{log}({\text{exp}(\text{log}(Lq_l)) + \text{exp}(\text{log}(Mp^{\star}_l / \hat{p}_{BS,t}))})\big)\Big) \\&- \Big[-\text{log}(M) + \text{log}\Big(\sum_{m=1}^M  \text{exp}\big(\text{log}({q_m})-\text{log}({\text{exp}(\text{log}(Lq_m)) + \text{exp}(\text{log}(Mp^{\star}_m / \hat{p}_{BS,t}))})\big)\Big)\Big]\\[8pt]
 &= -\text{log}(L) + \text{log}\Big(\sum_{l=1}^L \text{exp}\big(\text{log}({p^{\star}_l})-\text{log}({\text{exp}(\text{log}(L) + \text{log}(q_l))) + \text{exp}(\text{log}(M) + \text{log}(p^{\star}_l) - \text{log}(\hat{p}_{BS,t})))})\big)\Big) \\&- \Big[-\text{log}(M) + \underbrace{\text{log}\Big(\sum_{m=1}^M  \text{exp}\big(\text{log}({q_m})-\underbrace{\text{log}(\text{exp}(\text{log}(L) + \text{log}(q_m))) + \text{exp}(\text{log}(M) + \text{log}(p^{\star}_m) - \text{log}(\hat{p}_{BS,t}))}_{\text{Inner LogSumExp}}))\big)\Big)}_{\text{Outer LogSumExp}}\Big]\\
  \end{aligned}
  \end{equation*}
  
where the evaluated log likelihoods and the LogSumExp function defined above can be used to generate estimates of the logarithm of the marginal likelihood that are reasonably robust to numeric under- and overflow.  
\clearpage
\end{landscape}
\onecolumn
\section{Simulation Study Results}
\vspace{-0.5cm}
\label{simstudy}
% latex table generated in R 3.4.1 by xtable 1.8-2 package
% Tue Mar 06 11:11:31 2018
\begin{table*}[!h]
\centering
\begin{threeparttable}
 \scriptsize
        \setlength\tabcolsep{4pt}
        
        \renewcommand\arraystretch{0.4}
        \caption{Simulation Study Results for N=3000}

\begin{tabular}{cd{3.3}d{3.3}d{3.3}d{3.3}d{3.3}d{3.3}d{3.3}cr}
  \toprule
Parameter & \thead{True} &  \thead{Standard Prior} &  & \thead{SSVS} &  & \thead{Normal Gamma} &  \\ 
\cmidrule(l{3pt}r{3pt}){3-4}\cmidrule(l{3pt}r{3pt}){5-6}\cmidrule(l{3pt}r{3pt}){7-8}
   &  & \mc{Est.} & \mc{SE} & \mc{Est.} & \mc{SE} & \mc{Est.} & \mc{SE} \\ 
  \textbf{Group 1} &&&&&&&&&\\
  $\beta_{1,1}$ &  0.80 &   0.82 & 0.08 &   0.77 & 0.08 &   0.76 & 0.08 \\ 
  $\beta_{1,2}$ &  1.00 &   1.02 & 0.08 &   0.99 & 0.07 &   0.98 & 0.07 \\ 
  $\beta_{1,3}$ &  2.00 &   2.04 & 0.10 &   2.00 & 0.09 &   1.99 & 0.09 \\ 
  $\beta_{1,4}$ &  0.50 &   0.53 & 0.08 &   0.52 & 0.07 &   0.51 & 0.07 \\ 
  $\beta_{1,5}$ &  0.00 &  -0.01 & 0.08 &   0.00 & 0.06 &  -0.01 & 0.05 \\ 
  $\beta_{1,6}$ &  0.00 &  -0.01 & 0.08 &  -0.01 & 0.06 &   0.00 & 0.04 \\ 
  $\beta_{1,7}$ &  0.00 &  -0.01 & 0.08 &   0.00 & 0.06 &   0.00 & 0.05 \\ 
  $\beta_{1,8}$ &  0.00 &   0.00 & 0.07 &   0.00 & 0.06 &   0.00 & 0.04 \\ 
  $\beta_{1,9}$ &  0.00 &  -0.02 & 0.07 &  -0.01 & 0.06 &   0.00 & 0.04 \\ 
  $\beta_{1,10}$ &  0.00 &   0.03 & 0.07 &   0.02 & 0.06 &   0.01 & 0.04 \\ 
  $\beta_{1,11}$ &  0.00 &   0.02 & 0.07 &   0.01 & 0.06 &   0.00 & 0.04 \\ 
  $\beta_{1,12}$ &  0.00 &   0.00 & 0.07 &   0.00 & 0.06 &   0.00 & 0.04 \\ 
  $\beta_{1,13}$ &  0.00 &   0.00 & 0.07 &   0.00 & 0.06 &   0.00 & 0.04 \\ 
  $\beta_{1,14}$ &  0.00 &   0.00 & 0.07 &   0.00 & 0.06 &   0.00 & 0.03 \\ 
  $\beta_{1,15}$ &  0.00 &   0.01 & 0.07 &   0.01 & 0.06 &   0.00 & 0.04 \\ 
  $\beta_{1,16}$ &  0.00 &   0.01 & 0.07 &   0.00 & 0.06 &   0.00 & 0.04 \\ 
  $\beta_{1,17}$ &  0.00 &  -0.02 & 0.07 &  -0.01 & 0.06 &  -0.01 & 0.04 \\ 
  $\beta_{1,18}$ &  0.00 &   0.02 & 0.07 &   0.01 & 0.06 &   0.01 & 0.04 \\ 
  $\beta_{1,19}$ &  0.00 &   0.01 & 0.07 &   0.01 & 0.06 &   0.01 & 0.04 \\ 
  $\beta_{1,20}$ &  0.00 &  -0.04 & 0.07 &  -0.03 & 0.06 &  -0.02 & 0.04 \\ 
  \midrule
    \textbf{Group 2} &&&&&&&&&\\
  $\beta_{2,1}$ &  0.30 &   0.28 & 0.08 &   0.22 & 0.08 &   0.21 & 0.09 \\ 
  $\beta_{2,2}$ &  0.00 &   0.00 & 0.08 &   0.00 & 0.06 &   0.00 & 0.05 \\ 
  $\beta_{2,3}$ &  0.00 &  -0.03 & 0.10 &  -0.02 & 0.08 &  -0.01 & 0.06 \\ 
  $\beta_{2,4}$ &  0.00 &   0.00 & 0.08 &   0.00 & 0.06 &   0.00 & 0.04 \\ 
  $\beta_{2,5}$ & -1.00 &  -1.02 & 0.08 &  -1.00 & 0.08 &  -0.99 & 0.07 \\ 
  $\beta_{2,6}$ &  1.70 &   1.73 & 0.09 &   1.69 & 0.08 &   1.69 & 0.08 \\ 
  $\beta_{2,7}$ & -2.00 &  -2.06 & 0.10 &  -2.01 & 0.09 &  -2.00 & 0.08 \\ 
  $\beta_{2,8}$ &  0.00 &   0.00 & 0.08 &   0.00 & 0.06 &   0.00 & 0.04 \\ 
  $\beta_{2,9}$ &  0.00 &  -0.02 & 0.08 &  -0.01 & 0.06 &   0.00 & 0.03 \\ 
  $\beta_{2,10}$ &  0.00 &  -0.01 & 0.08 &  -0.01 & 0.06 &  -0.01 & 0.04 \\ 
  $\beta_{2,11}$ &  0.00 &   0.00 & 0.08 &   0.00 & 0.06 &  -0.01 & 0.04 \\ 
  $\beta_{2,12}$ &  0.00 &   0.01 & 0.08 &   0.01 & 0.06 &   0.00 & 0.03 \\ 
  $\beta_{2,13}$ &  0.00 &   0.02 & 0.08 &   0.02 & 0.06 &   0.01 & 0.04 \\ 
  $\beta_{2,14}$ &  0.00 &   0.01 & 0.08 &   0.01 & 0.06 &   0.01 & 0.04 \\ 
  $\beta_{2,15}$ &  0.00 &   0.00 & 0.08 &   0.00 & 0.06 &   0.00 & 0.04 \\ 
  $\beta_{2,16}$ &  0.00 &   0.01 & 0.08 &   0.00 & 0.06 &   0.00 & 0.04 \\ 
  $\beta_{2,17}$ &  0.00 &   0.00 & 0.08 &   0.00 & 0.06 &   0.00 & 0.04 \\ 
  $\beta_{2,18}$ &  0.00 &   0.01 & 0.08 &   0.01 & 0.06 &   0.00 & 0.04 \\ 
  $\beta_{2,19}$ &  0.00 &   0.01 & 0.08 &   0.00 & 0.06 &   0.00 & 0.04 \\ 
  $\beta_{2,20}$ &  0.00 &   0.00 & 0.08 &   0.00 & 0.06 &   0.01 & 0.04 \\ 
  \midrule
      \textbf{Group 3} &&&&&&&&&\\
  $\beta_{3,1}$ &  0.30 &   0.30 & 0.07 &   0.24 & 0.08 &   0.24 & 0.08 \\ 
  $\beta_{3,2}$ &  1.00 &   1.04 & 0.08 &   1.02 & 0.08 &   1.01 & 0.07 \\ 
  $\beta_{3,3}$ & -2.00 &  -2.06 & 0.11 &  -2.01 & 0.10 &  -1.99 & 0.09 \\ 
  $\beta_{3,4}$ &  0.80 &   0.82 & 0.08 &   0.80 & 0.07 &   0.79 & 0.07 \\ 
  $\beta_{3,5}$ &  0.90 &   0.94 & 0.08 &   0.92 & 0.08 &   0.91 & 0.07 \\ 
  $\beta_{3,6}$ &  0.00 &   0.00 & 0.08 &   0.00 & 0.06 &   0.00 & 0.04 \\ 
  $\beta_{3,7}$ &  0.00 &  -0.03 & 0.08 &  -0.02 & 0.06 &  -0.01 & 0.04 \\ 
  $\beta_{3,8}$ &  0.00 &   0.01 & 0.07 &   0.01 & 0.06 &   0.01 & 0.03 \\ 
  $\beta_{3,9}$ &  0.00 &  -0.03 & 0.07 &  -0.01 & 0.06 &   0.00 & 0.04 \\ 
  $\beta_{3,10}$ &  0.00 &   0.00 & 0.07 &   0.00 & 0.06 &   0.00 & 0.04 \\ 
  $\beta_{3,11}$ &  0.00 &   0.03 & 0.07 &   0.02 & 0.06 &   0.01 & 0.05 \\ 
  $\beta_{3,12}$ &  0.00 &  -0.01 & 0.07 &  -0.01 & 0.06 &  -0.01 & 0.04 \\ 
  $\beta_{3,13}$ &  0.00 &   0.01 & 0.07 &   0.01 & 0.06 &   0.00 & 0.04 \\ 
  $\beta_{3,14}$ &  0.00 &   0.00 & 0.07 &   0.00 & 0.06 &   0.00 & 0.04 \\ 
  $\beta_{3,15}$ &  0.00 &   0.03 & 0.07 &   0.02 & 0.06 &   0.01 & 0.04 \\ 
  $\beta_{3,16}$ &  0.00 &   0.01 & 0.07 &   0.00 & 0.06 &   0.00 & 0.04 \\ 
  $\beta_{3,17}$ &  0.00 &   0.00 & 0.07 &   0.00 & 0.06 &   0.00 & 0.04 \\ 
  $\beta_{3,18}$ &  0.00 &   0.01 & 0.07 &   0.00 & 0.06 &   0.00 & 0.04 \\ 
  $\beta_{3,19}$ &  0.00 &   0.00 & 0.07 &   0.00 & 0.06 &   0.00 & 0.04 \\ 
  $\beta_{3,20}$ &  0.00 &   0.00 & 0.07 &   0.00 & 0.06 &   0.00 & 0.04 \\ 
  \midrule
  RMSE (Zeroes) & - &   6.16 &  &   3.94 &  &   2.16 &  \\ 
  RMSE (Nonzeroes) & - &   8.83 &  &   8.35 &  &   8.35 &  \\ 
  RMSE (Overall) & - &   6.69 &  &   4.94 &  &   3.92 &  \\ 
  RMSE (Predicted Probabilities) & - &   3.60 &  &   2.94 &  &   2.29 &  \\ 
  Time in sec. & - & 448.56 &  & 449.60 &  & 461.41 &  \\ 
  \bottomrule

\end{tabular}
\begin{tablenotes}
\item The estimates correspond to the mean of the posterior means across 25 runs. RMSEs are multiplied by a factor of 100.
\end{tablenotes}
\end{threeparttable}
\end{table*}
\clearpage

\begin{table*}[!h]
\centering
\begin{threeparttable}
 \scriptsize
        \setlength\tabcolsep{4pt}
        
        \renewcommand\arraystretch{0.4}
        \caption{Simulation Study Results for N=300}

\begin{tabular}{cd{3.3}d{3.3}d{3.3}d{3.3}d{3.3}d{3.3}d{3.3}cr}
  \toprule
Parameter & \thead{True} &  \thead{Standard Prior} &  & \thead{SSVS} &  & \thead{Normal Gamma} &  \\ 
\cmidrule(l{3pt}r{3pt}){3-4}\cmidrule(l{3pt}r{3pt}){5-6}\cmidrule(l{3pt}r{3pt}){7-8}
   &  & \mc{Est.} & \mc{SE} & \mc{Est.} & \mc{SE} & \mc{Est.} & \mc{SE} \\ 
  \textbf{Group 1} &&&&&&&&&\\
  $\beta_{1,1}$ &  0.80 &  1.08 & 0.30 &  0.60 & 0.24 &  0.52 & 0.23 \\ 
  $\beta_{1,2}$ &  1.00 &  1.42 & 0.33 &  0.99 & 0.25 &  0.92 & 0.27 \\ 
  $\beta_{1,3}$ &  2.00 &  2.66 & 0.42 &  1.96 & 0.29 &  1.97 & 0.30 \\ 
  $\beta_{1,4}$ &  0.50 &  0.66 & 0.29 &  0.35 & 0.23 &  0.28 & 0.21 \\ 
  $\beta_{1,5}$ &  0.00 &  0.04 & 0.32 &  0.01 & 0.19 & -0.01 & 0.16 \\ 
  $\beta_{1,6}$ &  0.00 &  0.01 & 0.31 & -0.03 & 0.16 & -0.01 & 0.13 \\ 
  $\beta_{1,7}$ &  0.00 &  0.06 & 0.32 &  0.02 & 0.17 &  0.00 & 0.13 \\ 
  $\beta_{1,8}$ &  0.00 & -0.06 & 0.29 &  0.00 & 0.14 &  0.00 & 0.11 \\ 
  $\beta_{1,9}$ &  0.00 & -0.01 & 0.29 & -0.01 & 0.15 & -0.01 & 0.12 \\ 
  $\beta_{1,10}$ &  0.00 &  0.06 & 0.29 &  0.03 & 0.15 &  0.02 & 0.12 \\ 
  $\beta_{1,11}$ &  0.00 &  0.00 & 0.30 & -0.01 & 0.15 & -0.01 & 0.11 \\ 
  $\beta_{1,12}$ &  0.00 &  0.00 & 0.29 & -0.01 & 0.16 & -0.01 & 0.13 \\ 
  $\beta_{1,13}$ &  0.00 & -0.06 & 0.29 &  0.00 & 0.16 &  0.00 & 0.13 \\ 
  $\beta_{1,14}$ &  0.00 &  0.03 & 0.29 &  0.02 & 0.14 &  0.01 & 0.11 \\ 
  $\beta_{1,15}$ &  0.00 & -0.17 & 0.30 & -0.07 & 0.16 & -0.06 & 0.13 \\ 
  $\beta_{1,16}$ &  0.00 &  0.05 & 0.29 &  0.03 & 0.14 &  0.02 & 0.11 \\ 
  $\beta_{1,17}$ &  0.00 &  0.00 & 0.29 & -0.02 & 0.14 & -0.01 & 0.11 \\ 
  $\beta_{1,18}$ &  0.00 &  0.03 & 0.29 &  0.01 & 0.14 &  0.00 & 0.11 \\ 
  $\beta_{1,19}$ &  0.00 &  0.02 & 0.29 &  0.00 & 0.14 &  0.00 & 0.11 \\ 
  $\beta_{1,20}$ &  0.00 &  0.11 & 0.29 &  0.03 & 0.15 &  0.02 & 0.12 \\ 
  \midrule
    \textbf{Group 2} &&&&&&&&&\\
  $\beta_{2,1}$ &  0.30 &  0.37 & 0.30 &  0.08 & 0.18 &  0.04 & 0.15 \\ 
  $\beta_{2,2}$ &  0.00 & -0.03 & 0.31 & -0.05 & 0.18 & -0.07 & 0.17 \\ 
  $\beta_{2,3}$ &  0.00 & -0.18 & 0.41 & -0.06 & 0.23 & -0.06 & 0.21 \\ 
  $\beta_{2,4}$ &  0.00 & -0.12 & 0.31 & -0.12 & 0.19 & -0.13 & 0.18 \\ 
  $\beta_{2,5}$ & -1.00 & -1.41 & 0.33 & -1.03 & 0.26 & -1.04 & 0.27 \\ 
  $\beta_{2,6}$ &  1.70 &  2.43 & 0.37 &  1.69 & 0.25 &  1.72 & 0.26 \\ 
  $\beta_{2,7}$ & -2.00 & -2.85 & 0.40 & -2.03 & 0.27 & -2.09 & 0.28 \\ 
  $\beta_{2,8}$ &  0.00 & -0.10 & 0.30 & -0.01 & 0.15 &  0.00 & 0.12 \\ 
  $\beta_{2,9}$ &  0.00 &  0.13 & 0.31 &  0.04 & 0.16 &  0.03 & 0.13 \\ 
  $\beta_{2,10}$ &  0.00 & -0.03 & 0.31 & -0.03 & 0.16 & -0.02 & 0.13 \\ 
  $\beta_{2,11}$ &  0.00 &  0.06 & 0.31 &  0.02 & 0.16 &  0.01 & 0.13 \\ 
  $\beta_{2,12}$ &  0.00 & -0.03 & 0.30 & -0.03 & 0.16 & -0.02 & 0.14 \\ 
  $\beta_{2,13}$ &  0.00 & -0.07 & 0.30 & -0.01 & 0.15 &  0.00 & 0.13 \\ 
  $\beta_{2,14}$ &  0.00 & -0.06 & 0.31 & -0.02 & 0.15 & -0.02 & 0.12 \\ 
  $\beta_{2,15}$ &  0.00 &  0.00 & 0.30 &  0.01 & 0.15 &  0.01 & 0.12 \\ 
  $\beta_{2,16}$ &  0.00 & -0.06 & 0.30 & -0.02 & 0.15 & -0.02 & 0.12 \\ 
  $\beta_{2,17}$ &  0.00 &  0.06 & 0.31 &  0.01 & 0.16 &  0.01 & 0.14 \\ 
  $\beta_{2,18}$ &  0.00 & -0.10 & 0.31 & -0.03 & 0.15 & -0.02 & 0.12 \\ 
  $\beta_{2,19}$ &  0.00 &  0.08 & 0.30 &  0.03 & 0.15 &  0.02 & 0.12 \\ 
  $\beta_{2,20}$ &  0.00 &  0.00 & 0.31 & -0.02 & 0.16 & -0.02 & 0.13 \\ 
  \midrule
      \textbf{Group 3} &&&&&&&&&\\
  $\beta_{3,1}$ &  0.30 &  0.44 & 0.30 &  0.11 & 0.18 &  0.06 & 0.15 \\ 
  $\beta_{3,2}$ &  1.00 &  1.43 & 0.33 &  0.98 & 0.26 &  0.91 & 0.27 \\ 
  $\beta_{3,3}$ & -2.00 & -2.92 & 0.45 & -2.04 & 0.30 & -2.07 & 0.32 \\ 
  $\beta_{3,4}$ &  0.80 &  0.98 & 0.31 &  0.60 & 0.26 &  0.52 & 0.25 \\ 
  $\beta_{3,5}$ &  0.90 &  1.28 & 0.33 &  0.88 & 0.26 &  0.82 & 0.27 \\ 
  $\beta_{3,6}$ &  0.00 &  0.10 & 0.31 &  0.02 & 0.16 &  0.03 & 0.13 \\ 
  $\beta_{3,7}$ &  0.00 &  0.11 & 0.32 &  0.03 & 0.16 &  0.00 & 0.13 \\ 
  $\beta_{3,8}$ &  0.00 & -0.05 & 0.30 &  0.00 & 0.15 &  0.00 & 0.12 \\ 
  $\beta_{3,9}$ &  0.00 &  0.02 & 0.30 &  0.01 & 0.17 &  0.01 & 0.14 \\ 
  $\beta_{3,10}$ &  0.00 & -0.04 & 0.30 & -0.03 & 0.14 & -0.02 & 0.11 \\ 
  $\beta_{3,11}$ &  0.00 & -0.01 & 0.30 & -0.01 & 0.15 &  0.00 & 0.12 \\ 
  $\beta_{3,12}$ &  0.00 &  0.07 & 0.30 &  0.02 & 0.15 &  0.01 & 0.13 \\ 
  $\beta_{3,13}$ &  0.00 & -0.06 & 0.30 & -0.02 & 0.16 & -0.02 & 0.13 \\ 
  $\beta_{3,14}$ &  0.00 &  0.05 & 0.30 &  0.03 & 0.16 &  0.02 & 0.13 \\ 
  $\beta_{3,15}$ &  0.00 &  0.02 & 0.30 &  0.03 & 0.15 &  0.03 & 0.12 \\ 
  $\beta_{3,16}$ &  0.00 & -0.07 & 0.30 & -0.02 & 0.14 & -0.01 & 0.11 \\ 
  $\beta_{3,17}$ &  0.00 &  0.00 & 0.30 & -0.01 & 0.15 & -0.01 & 0.12 \\ 
  $\beta_{3,18}$ &  0.00 &  0.02 & 0.29 &  0.01 & 0.15 &  0.01 & 0.12 \\ 
  $\beta_{3,19}$ &  0.00 &  0.05 & 0.30 & -0.01 & 0.15 & -0.01 & 0.13 \\ 
  $\beta_{3,20}$ &  0.00 &  0.10 & 0.30 &  0.03 & 0.16 &  0.02 & 0.14 \\ 
  \midrule
  RMSE (Zeroes) & - & 29.19 &  &  9.62 &  &  7.51 &  \\ 
  RMSE (Nonzeroes) & - & 63.39 &  & 26.55 &  & 30.57 &  \\ 
  RMSE (Overall) & - & 37.31 &  & 13.94 &  & 14.19 &  \\ 
  RMSE (Predicted Probabilities) & - & 12.94 &  &  7.99 &  &  7.55 &  \\ 
  Time in sec. & - & 57.30 &  & 70.80 &  & 85.50 &  \\ 
  \bottomrule

\end{tabular}
\begin{tablenotes}
\item The estimates correspond to the mean of the posterior means across 25 runs. RMSEs are multiplied by a factor of 100.
\end{tablenotes}
\end{threeparttable}
\end{table*}

\begin{table*}
\centering
\begin{threeparttable}
 \scriptsize
        \setlength\tabcolsep{4pt}
        
        \renewcommand\arraystretch{0.4}
        \caption{Simulation Study Results for N=100}

\begin{tabular}{cd{3.3}d{3.3}d{3.3}d{3.3}d{3.3}d{3.3}d{3.3}cr}
  \toprule
Parameter & \thead{True} &  \thead{Standard Prior} &  & \thead{SSVS} &  & \thead{Normal Gamma} &  \\ 
\cmidrule(l{3pt}r{3pt}){3-4}\cmidrule(l{3pt}r{3pt}){5-6}\cmidrule(l{3pt}r{3pt}){7-8}
   &  & \mc{Est.} & \mc{SE} & \mc{Est.} & \mc{SE} & \mc{Est.} & \mc{SE} \\ 
  \textbf{Group 1} &&&&&&&&&\\
  $\beta_{1,1}$ &  0.80 &   2.50 & 1.05 &  0.52 & 0.39 &  0.46 & 0.39 \\ 
  $\beta_{1,2}$ &  1.00 &   2.93 & 0.99 &  0.73 & 0.45 &  0.60 & 0.49 \\ 
  $\beta_{1,3}$ &  2.00 &   6.45 & 1.33 &  2.04 & 0.49 &  2.31 & 0.71 \\ 
  $\beta_{1,4}$ &  0.50 &   1.27 & 0.98 &  0.17 & 0.35 &  0.08 & 0.31 \\ 
  $\beta_{1,5}$ &  0.00 &   0.35 & 1.03 &  0.06 & 0.36 &  0.00 & 0.32 \\ 
  $\beta_{1,6}$ &  0.00 &  -0.62 & 1.05 & -0.15 & 0.34 & -0.07 & 0.27 \\ 
  $\beta_{1,7}$ &  0.00 &   0.58 & 1.05 &  0.10 & 0.31 &  0.04 & 0.26 \\ 
  $\beta_{1,8}$ &  0.00 &   0.14 & 0.98 &  0.02 & 0.29 &  0.01 & 0.24 \\ 
  $\beta_{1,9}$ &  0.00 &   0.20 & 0.97 &  0.02 & 0.28 &  0.01 & 0.23 \\ 
  $\beta_{1,10}$ &  0.00 &   0.14 & 0.99 &  0.02 & 0.30 &  0.00 & 0.25 \\ 
  $\beta_{1,11}$ &  0.00 &  -0.32 & 0.98 & -0.03 & 0.28 & -0.01 & 0.22 \\ 
  $\beta_{1,12}$ &  0.00 &  -0.46 & 0.98 & -0.09 & 0.29 & -0.07 & 0.24 \\ 
  $\beta_{1,13}$ &  0.00 &   0.08 & 0.99 &  0.01 & 0.30 &  0.00 & 0.26 \\ 
  $\beta_{1,14}$ &  0.00 &  -0.31 & 1.00 & -0.05 & 0.31 & -0.06 & 0.26 \\ 
  $\beta_{1,15}$ &  0.00 &  -0.13 & 1.01 & -0.01 & 0.29 &  0.00 & 0.23 \\ 
  $\beta_{1,16}$ &  0.00 &   0.21 & 0.93 &  0.00 & 0.30 &  0.00 & 0.26 \\ 
  $\beta_{1,17}$ &  0.00 &  -0.21 & 0.98 & -0.03 & 0.30 & -0.05 & 0.24 \\ 
  $\beta_{1,18}$ &  0.00 &  -0.23 & 0.98 &  0.00 & 0.29 &  0.00 & 0.24 \\ 
  $\beta_{1,19}$ &  0.00 &  -0.28 & 1.01 & -0.02 & 0.30 & -0.01 & 0.24 \\ 
  $\beta_{1,20}$ &  0.00 &   0.14 & 1.02 &  0.03 & 0.30 &  0.00 & 0.26 \\
  \midrule
    \textbf{Group 2} &&&&&&&&&\\
  $\beta_{2,1}$ &  0.30 &   0.21 & 0.97 & -0.03 & 0.28 & -0.04 & 0.23 \\ 
  $\beta_{2,2}$ &  0.00 &  -0.32 & 0.95 & -0.21 & 0.34 & -0.26 & 0.35 \\ 
  $\beta_{2,3}$ &  0.00 &   0.08 & 1.18 &  0.00 & 0.36 & -0.06 & 0.42 \\ 
  $\beta_{2,4}$ &  0.00 &  -0.20 & 0.97 & -0.15 & 0.33 & -0.21 & 0.33 \\ 
  $\beta_{2,5}$ & -1.00 &  -3.14 & 1.07 & -0.95 & 0.48 & -1.09 & 0.60 \\ 
  $\beta_{2,6}$ &  1.70 &   4.73 & 1.09 &  1.55 & 0.44 &  1.84 & 0.58 \\ 
  $\beta_{2,7}$ & -2.00 &  -5.58 & 1.19 & -1.79 & 0.45 & -2.12 & 0.63 \\ 
  $\beta_{2,8}$ &  0.00 &   0.28 & 0.95 &  0.02 & 0.27 &  0.02 & 0.23 \\ 
  $\beta_{2,9}$ &  0.00 &   0.40 & 0.95 &  0.03 & 0.26 &  0.02 & 0.22 \\ 
  $\beta_{2,10}$ &  0.00 &   0.49 & 0.96 &  0.05 & 0.28 &  0.04 & 0.23 \\ 
  $\beta_{2,11}$ &  0.00 &  -0.13 & 0.93 &  0.00 & 0.25 &  0.00 & 0.20 \\ 
  $\beta_{2,12}$ &  0.00 &  -0.23 & 0.94 & -0.02 & 0.32 & -0.03 & 0.29 \\ 
  $\beta_{2,13}$ &  0.00 &  -0.04 & 0.91 & -0.01 & 0.27 & -0.01 & 0.23 \\ 
  $\beta_{2,14}$ &  0.00 &  -0.32 & 0.96 & -0.03 & 0.29 & -0.04 & 0.25 \\ 
  $\beta_{2,15}$ &  0.00 &  -0.03 & 0.94 &  0.02 & 0.28 &  0.00 & 0.24 \\ 
  $\beta_{2,16}$ &  0.00 &   0.16 & 0.91 &  0.05 & 0.29 &  0.06 & 0.26 \\ 
  $\beta_{2,17}$ &  0.00 &  -0.28 & 0.93 & -0.04 & 0.28 & -0.04 & 0.24 \\ 
  $\beta_{2,18}$ &  0.00 &   0.00 & 0.97 & -0.01 & 0.27 & -0.01 & 0.23 \\ 
  $\beta_{2,19}$ &  0.00 &  -0.10 & 0.96 & -0.03 & 0.29 & -0.04 & 0.24 \\ 
  $\beta_{2,20}$ &  0.00 &  -0.19 & 0.98 & -0.01 & 0.29 & -0.04 & 0.26 \\ 
  \midrule
      \textbf{Group 3} &&&&&&&&&\\
  $\beta_{3,1}$ &  0.30 &   0.67 & 0.96 &  0.04 & 0.28 &  0.02 & 0.22 \\ 
  $\beta_{3,2}$ &  1.00 &   3.04 & 1.01 &  0.68 & 0.44 &  0.59 & 0.47 \\ 
  $\beta_{3,3}$ & -2.00 &  -5.89 & 1.28 & -1.90 & 0.48 & -2.17 & 0.72 \\ 
  $\beta_{3,4}$ &  0.80 &   2.57 & 1.02 &  0.55 & 0.44 &  0.44 & 0.43 \\ 
  $\beta_{3,5}$ &  0.90 &   3.40 & 1.08 &  0.91 & 0.49 &  0.79 & 0.54 \\ 
  $\beta_{3,6}$ &  0.00 &  -0.62 & 1.00 & -0.10 & 0.33 & -0.02 & 0.30 \\ 
  $\beta_{3,7}$ &  0.00 &   0.72 & 1.00 &  0.15 & 0.34 &  0.09 & 0.29 \\ 
  $\beta_{3,8}$ &  0.00 &  -0.19 & 0.98 & -0.06 & 0.31 & -0.05 & 0.25 \\ 
  $\beta_{3,9}$ &  0.00 &   0.25 & 0.94 &  0.02 & 0.28 &  0.03 & 0.23 \\ 
  $\beta_{3,10}$ &  0.00 &  -0.42 & 0.96 & -0.09 & 0.31 & -0.08 & 0.26 \\ 
  $\beta_{3,11}$ &  0.00 &  -0.38 & 0.93 & -0.03 & 0.27 & -0.02 & 0.22 \\ 
  $\beta_{3,12}$ &  0.00 &  -0.15 & 0.95 & -0.04 & 0.32 & -0.03 & 0.27 \\ 
  $\beta_{3,13}$ &  0.00 &  -0.04 & 0.93 & -0.03 & 0.29 & -0.01 & 0.24 \\ 
  $\beta_{3,14}$ &  0.00 &   0.22 & 0.97 &  0.08 & 0.30 &  0.04 & 0.26 \\ 
  $\beta_{3,15}$ &  0.00 &   0.00 & 0.96 &  0.06 & 0.28 &  0.03 & 0.22 \\ 
  $\beta_{3,16}$ &  0.00 &  -0.19 & 0.90 & -0.05 & 0.29 & -0.04 & 0.24 \\ 
  $\beta_{3,17}$ &  0.00 &  -0.32 & 0.97 &  0.00 & 0.30 &  0.00 & 0.24 \\ 
  $\beta_{3,18}$ &  0.00 &  -0.08 & 0.98 &  0.00 & 0.31 &  0.00 & 0.26 \\ 
  $\beta_{3,19}$ &  0.00 &  -0.12 & 0.99 &  0.00 & 0.31 &  0.00 & 0.26 \\ 
  $\beta_{3,20}$ &  0.00 &  -0.40 & 0.97 & -0.04 & 0.28 & -0.06 & 0.24 \\
  \midrule
  RMSE (Zeroes) & - & 119.16 &  & 18.90 &  & 16.20 &  \\ 
  RMSE (Nonzeroes) & - & 286.90 &  & 41.97 &  & 60.93 &  \\ 
  RMSE (Overall) & - & 159.50 &  & 24.39 &  & 28.87 &  \\ 
  RMSE (Predicted Probabilities) & - &  23.29 &  & 13.46 &  & 12.96 &  \\ 
  Time in sec. & - &  28.74 &  & 41.81 &  & 56.32 &  \\ 
  \bottomrule

\end{tabular}
\begin{tablenotes}
\item The estimates correspond to the mean of the posterior means across 25 runs. RMSEs are multiplied by a factor of 100.
\end{tablenotes}
\end{threeparttable}
\end{table*}

%\begin{acknowledgements}
%\end{acknowledgements}

\end{document}